\documentclass[reqno,centertags,11pt]{amsart}
\usepackage{amsmath,amsthm,amscd,amssymb}
\usepackage{latexsym}
\usepackage{graphicx,psfrag}
\usepackage{pstricks}
\usepackage{pst-node}

\usepackage {verbatim}

\newcommand{\N}{{\mathbb{N}}}

\newcommand{\R}{{\mathbb{R}}}

\newcommand{\C}{{\mathbb{C}}}

\newcommand{\Z}{{\mathbb{Z}}}

\newcommand{\ol}{\overline}

\newcommand{\wti}{\widetilde  }

\newcommand{\beq}{\begin{equation}}
\newcommand{\eeq}{\end{equation}}
\newcommand{\bdm}{\begin{displaymath}}
\newcommand{\edm}{\end{displaymath}}
\newcommand{\ba}{\begin{align}}
\newcommand{\ea}{\end{align}}
\newcommand{\bpf}{\begin{proof}}
\newcommand{\epf}{\end{proof}}

\newcommand{\la}{\langle}
\newcommand{\ra}{\rangle}

\newcommand{\supp}{\mathrm{supp}\, }               
\newcommand{\dist}{\mathrm{dist}}               

\newcommand{\veps}{\varepsilon}

\newcommand{\re}{\mathrm{Re}}
\newcommand{\im}{\mathrm{Im}}

\newcommand{\id}{\mathbf{1}}                

\allowdisplaybreaks

\newcommand{\calQ}{\mathcal{Q}}

\newtheorem{thm}{Theorem}
\newtheorem{prop}[thm]{Proposition}
\newtheorem{lem}[thm]{Lemma}
\newtheorem{cor}[thm]{Corollary}

\theoremstyle{definition}

\newtheorem{remark}[thm]{Remark}
\newtheorem{remarks}[thm]{Remarks}

\newcounter{theoremi}[thm]

\newcommand{\itemthm}{\refstepcounter{theoremi} {\rm(\roman{theoremi})}{~}}

\numberwithin{thm}{section}
\numberwithin{equation}{section}

\begin{document}

\title[Super-exponential decay of Diffraction Managed Solitons]{%
    Super-exponential decay of Diffraction Managed Solitons}
\author[D.~Hundertmark and Y.-R.~Lee]{Dirk Hundertmark and Young-Ran~Lee}
\address{Department of Mathematics, Altgeld Hall,
    and Institute for Condensed Matter Theory at the
    University of Illinois at Urbana--Champaign,
    1409 W.~Green Street, Urbana, IL 61801.}%
\email{dirk@math.uiuc.edu}%
\address{Department of Mathematical Sciences
KAIST (Korea Advanced Institute of Science and Technology),
335 Gwahangno, Yuseong-gu, Daejeon, 305-701, Republic of Korea.}%
\email{youngranlee@kaist.ac.kr}

\thanks{\copyright 2008 by the authors. Faithful reproduction of this article,
        in its entirety, by any means is permitted for non-commercial purposes}
\date{16 April 2008, version 7-3, file diffms-7-3.tex}

\begin{abstract}
This is the second part of a series of papers where we develop rigorous decay estimates for
breather solutions of an averaged version of the non-linear Schr\"odinger equation. In this
part we study the diffraction managed discrete non-linear Schr\"odinger equation, an
equation which describes coupled waveguide arrays with periodic diffraction management
geometries.
We show that, for vanishing average diffraction, all solutions of the non-linear and non-local
diffraction management equation decay super-exponentially.
As a byproduct of our method, we also have a simple proof of existence of diffraction
managed solitons in the case of vanishing average diffraction.
\end{abstract}

\maketitle

\section{Introduction}\label{sec1}
Solitons, localized coherent structures resulting from a balance of non-linear and
dispersive effects, have been the focus of an intense research activity over the
last decades, see \cite{SCS00,TT01}. Besides solitons in the continuum, discrete
solitons have emerged in such diverse areas as solid states physics, some biological
systems, Bose-Einstein condensation, and in discrete non-linear optics, e.g., optical
waveguide arrays, \cite{CBG-J94,Davydov73,Scott85,SKES03,TS98}.
The model describing this range of phenomena is given by the discrete non--linear Schr\"odinger equation
 \beq\label{eq:discreteNLS-0}
    i \frac{\partial}{\partial \xi} u(x) + d(\xi) (\Delta u)(x) + |u(x)|^2 u(x) = 0
 \eeq
where for waveguide arrays $\xi$ is the distance along the waveguide, $x\in\Z$  the location
of the array element, $\Delta$ the discrete Laplacian given by
$(\Delta f)(x)= f(x+1)+ f(x-1) - 2f(x)$ for all $x\in \Z$, and $d(\xi)$ the
total diffraction along the waveguide.

Nearly a decade after their theoretical prediction in \cite{CJ88}, discrete solitons in an
optical waveguide array were studied experimentally and, as in the continuous case localized
stable non-linear waves where found \cite{ESMBA98}.
Similar to the continuous case, i.e, non-linear fiber-optics, where the dispersion management
technique introduced by \cite{LKC80} in 1980 turned out to be enormously successful in creating stable
low power pulses by periodically varying the dispersion along the glass--fiber cable, see
\cite{AB98,GT96a,GT96b,KH97,Kurtzge93,LK98,MGLWXK03,MM99,MMGNMG-NV99,TDNMSF99} and the
survey article \cite{Tetal03}, the diffraction management technique was proposed much more recently
in \cite{ESMA00} in order to create low power stable discrete pulses
by periodically varying the diffraction in discrete optical waveguide arrays. In this case, the total diffraction
$d(\xi)$ along the waveguide is given by
 \beq\label{eq:dispersion}
    d(\xi) = \veps^{-1} \wti{d}(\xi/\veps) + d_{av} .
 \eeq
Here $d_{av} \ge 0$ is the average component of the diffraction and $\wti{d}$ its
mean zero part. Note that unlike in the continuum case, the diffraction management technique
uses the \emph{geometry} of the waveguide to achieve a periodically varying diffraction, see
\cite{ESMA00}.

In the region of \emph{strong diffraction management} $\veps$ is a small positive
parameter. Rescaling $t=\xi/\veps$, \eqref{eq:discreteNLS-0} is equivalent to
 \beq\label{eq:discreteNLS}
    i \frac{\partial}{\partial t} u + \wti{d}(t) \Delta u
    + \veps\big(d_\text{av} \Delta u + |u|^2 u\big) = 0\, .
 \eeq

For small $\veps$ an average equation which describes the slow evolution of solutions of
\eqref{eq:discreteNLS} was derived and numerical studies showed that this average equation possesses
stable solutions which evolve nearly periodically when used as initial data in the
diffraction managed non-linear discrete Schr\"odinger equation, \cite{AM01,AM02,AM03}.
Normalizing the period in the fast variable $t$ to one, the average equation for the slow
part $v$ of solutions of \eqref{eq:discreteNLS} is given by
 \beq\label{eq:DiffMS-time}
   i \frac{\partial}{\partial t}v  +\veps d_{\text{av}}\Delta v + \veps Q(v,v,v) = 0
 \eeq
where
 \beq\label{eq:Q-1}
   Q(v_1,v_2,v_3):= \int_0^1 T_s^{-1}\big[ T_s v_1 \ol{T_sv_2} T_sv_3 \big]\, ds
 \eeq
with $T_s:= e^{i D(s)\Delta}$ and $D(s)= \int_0^s \wti{d}(\xi)\,d\xi$  the solution operator for the
free discrete Schr\"odinger equation with periodically varying diffraction
 \beq\label{eq:free}
    i \frac{\partial}{\partial t} v = - \wti{d}(t)\Delta v .
 \eeq
One should keep in mind that the variable $t$ denotes the distance along the waveguide.
Physically it makes sense to assume that the diffraction profile $\wti{d}$ is bounded, or
even piecewise constant along the waveguide. This assumption was made
in \cite{Moeser05,Panayotaros05,Stanislavova07}.
For our results we need only to assume that its integral $D$
is bounded over one period in the fast variable $t$,
 \beq\label{eq:D-bounded}
    \tau:= \sup_{t\in[0,1]}|D(t)|=\sup_{t\in[0,1]} \Big| \int_0^t \wti{d}(\xi)\, d\xi\Big|<\infty,
 \eeq
where we normalized, without loss of generality, the period in $t$ to one.

Using the same general method as in the continuum case, see, e.g., \cite{ZGJT01},
the averaged equation \eqref{eq:DiffMS-time} was derived
in \cite{AM01,AM02,AM03}, where it was expressed in the Fourier space.
The above formulation is from \cite{Moeser05, Panayotaros05}.
Note that the non-linear and non-local equation \eqref{eq:DiffMS-time} has an
associated (averaged) Hamiltonian given by
 \beq\label{eq:Hamiltonian}
    H(v) := \veps\big( \frac{d_{\text{av}}}{2} \la v,-\Delta v\ra
                        -\frac{1}{4} \calQ(v,v,v,v)
                \big)
 \eeq
with $\la g,f\ra:= \sum_{x\in\Z} \ol{g(x)} f(x)$ the usual scalar product on $l^2(\Z)$, which,
in our convention, is linear in the second component and anti-linear in the first and
 \beq\label{eq:calQ-1}
   \calQ(v_1,v_2,v_3,v_4):= \int_0^1 \sum_{x\in\Z} \ol{(T_sv_1)(x)}(T_sv_2)(x)\ol{(T_sv_3)(x)}(T_sv_4)(x)\, ds .
 \eeq

Following the procedure for the continuous case in \cite{ZGJT01}, it was shown in \cite{Moeser05}
that over long scales $0\le t\le C\veps^{-1}$ solutions of the non-autonomous equation
\eqref{eq:discreteNLS} stay $\veps$-close to solutions of the autonomous average equation
\eqref{eq:DiffMS-time} with the same initial condition.
Thus it is interesting to find stationary solutions of \eqref{eq:DiffMS-time}, which are precisely
the right initial conditions leading to breather-like nearly periodic solutions of
\eqref{eq:discreteNLS} on long scales $0\le t\le C\veps^{-1}$.
Making the ansatz $v(t,x)= e^{i\veps \omega t} f(x)$ in \eqref{eq:DiffMS-time} one arrives at the non-linear
and non-local eigenvalue problem
 \beq\label{eq:DiffMS-pos-d-av}
  -\omega f = -d_\text{av}\Delta f - Q(f,f,f).
 \eeq
Solutions of this equation can be found by minimizing the Hamiltonian  $H$ in \eqref{eq:Hamiltonian}
over functions $f\in l^2(\Z)$ with a fixed $l^2$-norm. The problem of constructing such minimizers
for positive average diffraction $d_\text{av}>0$ has been studied in \cite{Moeser05,Panayotaros05}
using a discrete version of Lions' concentration compactness method \cite{Lions84}. Moreover,
using by now classical arguments, see, \cite{Weinstein1,Weinstein2} or \cite{Cazenave03}, it was
noticed in \cite{Moeser05,Panayotaros05} that these minimizers are so-called \emph{orbitally stable},
explaining at least in part the strong stability properties of diffraction management.

Similar as in the continuous case, see \cite{Kunze04}, proving existence of minimizers for vanishing
average dispersion, $d_\text{av}=0$, i.e., existence of weak solutions $f\in l^2(\Z)$ of
 \beq\label{eq:DiffMS}
   \omega f =  Q(f,f,f)
 \eeq
is much harder and has only recently been established in \cite{Stanislavova07} using Ekeland's
variational principle, \cite{Ekeland72,Ekeland73}.
Moreover, it was shown in \cite{Stanislavova07}, that the corresponding
minimizer is decaying faster than polynomial, which again yields the orbital stability of
solutions of \eqref{eq:DiffMS-time} for $d_\text{av}=0$ and initial conditions close to a minimizer.

In this paper we continue our study of regularity properties of the dispersion management technique
initiated in \cite{100DMLee07} and study the decay properties of diffraction management solitons for
vanishing average dispersion, i.e., weak solutions of \eqref{eq:DiffMS}.
Our main result is a significant strengthening of the super-polynomial decay result for diffraction managed
solitons in \cite{Stanislavova07}.

\begin{thm}[Super-exponential decay]\label{thm:main-decay}
Assume that the diffraction profile obeys \eqref{eq:D-bounded}. Then any
weak solution $f\in l^2(\Z)$ of \eqref{eq:DiffMS} decays faster than any
exponential. More precisely, with $c=1+\ln(8\tau)$,
 \bdm
     |f(x)| \lesssim e^{-\frac{1}{4}(|x|+1)(\ln((|x|+1)/2)-c)} \quad \text{ for all }x\in\Z .
 \edm
\end{thm}

\begin{remarks}\itemthm In particular, for any $0<\mu<1/4$ Theorem \ref{thm:main-decay} yields
the bound
 \bdm
   |f(x)| \lesssim (|x|+1)^{-\mu(|x|+1)} \quad \text{ for all }x\in\Z.
 \edm
\itemthm The bound given in Theorem \ref{thm:main-decay} rigorously justifies the theoretical
and experimental conclusion of \cite{ESMA00}, that the diffraction management technique leads to
optical soliton like pulses along a waveguide array which are extremely well-localized along the
array elements. \\[0.2em]
\itemthm The super-exponential decay given in Theorem \ref{thm:main-decay} is in stark contrast
to the continuous case where one has, so far, only super-polynomial bounds on the decay of
dispersion management solitons, see \cite{100DMLee07}. It is believed that the decay in the
continuous case is exponential, see \cite{Lushnikov04} for convincing but non-rigorous
arguments.\\[0.2em]
\itemthm Weak solutions of \eqref{eq:DiffMS} are defined as
 \beq
   \omega \la g,f\ra = \la g, Q(f,f,f)\ra 
 \eeq
for any $g\in l^2(\Z)$. Recalling the definition \eqref{eq:calQ-1} for the four-linear
functional $\calQ$, a short calculation gives
 \beq
   \la f_1, Q(f_2,f_3,f_4)\ra  = \calQ(f_1,f_2,f_3,f_4)
 \eeq
 for any $f_j\in l^2(\Z)$, $j=1,2,3,4$. Thus $f$ is a weak solutions of \eqref{eq:DiffMS} if and only
 if
 \beq
   \omega \la g,f\ra = \calQ(g,f,f,f)
 \eeq
 for all $g\in l^2(\Z)$.\\[0.2em]
 \itemthm One easily sees that $\calQ(f,f,f,f)>0$ as soon as $f$ is not the zero function. Thus
 $\omega = \calQ(f,f,f,f)/\la f,f\ra>0$ for any non-zero solution of \eqref{eq:DiffMS}.
\end{remarks}

Our second result is a simple proof of existence of weak solutions of \eqref{eq:DiffMS} under
the weak condition \eqref{eq:D-bounded} on the diffraction profile.

\begin{thm}\label{thm:main-existence}
Assume that the diffraction profile obeys \eqref{eq:D-bounded}.
Let $\lambda>0$. There is an $f\in l^2(\Z)$ with $\|f\|_2^2=\lambda$
such that
 \bdm
   \calQ(f,f,f,f) = P_\lambda:= \sup \left\{ \calQ(g,g,g,g)\vert\, g\in l^2(\Z), \|g\|_2^2=\lambda \right\} .
 \edm
 This maximizer $f$ is also a weak solution of the diffraction management equation
 \eqref{eq:DiffMS} where $\omega>0 $ is a suitable Lagrange-multiplier.
\end{thm}

\begin{remark} The existence of a maximizer is non-trivial, even for the corresponding problem
with $d_\text{av}>0$, since the equation \eqref{eq:DiffMS}, respectively \eqref{eq:DiffMS-pos-d-av},
is invariant under translations and so is the corresponding energy functional $H$ given by
\eqref{eq:Hamiltonian}. Thus maximizing sequences for $\calQ$, respectively minimizing sequences for $H$, can
very easily converge to zero weakly. This was overcome using Lions' concentration compactness principle
in \cite{Moeser05,Panayotaros05} for positive average diffraction and, for vanishing average dispersion,
in \cite{Stanislavova07} using Ekeland's variational principle \cite{Ekeland72,Ekeland73,Jabri},
assuming that the diffraction profile is piecewise constant.
Besides holding under much more general conditions, our proof is rather direct and, we believe, simple.
We show that modulo translations any maximizing sequence has a strongly convergent subsequence, i.e.,
there is a sequence of shifts such that the shifted sequence, which by the translation invariance of the
problem is also a maximizing sequence, has a strongly convergent subsequence.
Our proof avoids the use of the concentration compactness principle
but relies instead on a discrete version of multi-linear Strichartz estimates, see Corollary
\ref{cor:multilinear} and Lemma \ref{lem:epsilon-fat-tail}.
\end{remark}

Our paper is organized as follows: In the next section we fix our notation and develop our
basic technical estimates, the discrete versions of bilinear and multi--linear
Strichartz estimates from Lemmata \ref{lem:strong-bilinear} and \ref{lem:strong-multilinear}
and Corollary \ref{cor:multilinear}. All results in Section \ref{sec:basic} are valid in
arbitrary dimension $d\ge 1$.
The proof of Theorem \ref{thm:main-decay} is given in Section \ref{sec:self-consistency},
see Theorem \ref{thm:super-exp-decay} and Corollary \ref{cor:super-exp-decay}.
Similar to our study of decay properties of dispersion managed solitons in \cite{100DMLee07},
the main tool in the proof of our super-exponential decay Theorem \ref{thm:main-decay} is
the self-consistency bound from Proposition \ref{prop:self-consistency} on the tail
distribution of weak solutions of the diffraction management equation \eqref{eq:DiffMS}.
Our existence proof for diffraction management solitons is given in Section \ref{sec:existence}.
It relies heavily on Lemma \ref{lem:epsilon-fat-tail}, which follows from the
enhanced multi--linear estimates of Corollary \ref{cor:multilinear}, and
a simple characterization of strong convergence in $l^2(\Z)$, or more generally, strong convergence
in $l^p(\Z^d)$ for $1\le p<\infty$, given in Lemma \ref{lem:compactness-criterion}.

\section{Basic estimates} \label{sec:basic}
In this section we consider $\Z^d$ for arbitrary dimension $d\ge 1$. First we introduce
some notation. By $\N$ we denote the natural numbers and $\N_0= \N\cup \{0\}$.
Given $n\in\N_0$, we denote $n!$ the factorial, $0!=1$ and $(n+1)!= (n+1)n!$.
The integers are denoted by $\Z$ and $\Z^d$ is the d-fold Euclidian product of $\Z$.
$l^p(\Z^d)$ is the usual sequence space with norm
 \beq
   \|f\|_p = \Big(\sum_{x\in\Z^d} |f(x)|^p  \Big)^{1/p}\quad \text{for } 1\le p<\infty
 \eeq
and
 \beq
   \|f\|_\infty = \sup_{x\in\Z^d}|f(x)| .
 \eeq
Of course, for $p=2$ we get the Hilbert space of square summable sequences indexed by $\Z^d$.
In this case we use
 \beq\label{eq:scalar-product}
    \la f,g \ra := \sum_{x\in\Z^d} \ol{f(x)} g(x)
 \eeq
for the scalar product on $l^2(\Z^d)$. Here $\ol{z}$ is the complex conjugate of a complex number $z$.
The real and imaginary parts of a complex number are given by
$\re(z)= \tfrac{1}{2}(z+\ol{z})$ and $\im(z)= \tfrac{1}{2i}(z-\ol{z})$.
Note that in our convention the scalar product given by \eqref{eq:scalar-product} is linear in
the second argument and anti-linear in the first. The discrete Laplacian on $\Z^d$ is given by
 \beq
   \Delta f(x) = \sum_{|\nu|=1} f(x+\nu) - 2d f(x)
 \eeq
where we take $|x|= \sum_{j=1}^d |x_j|$ for the norm on $\Z^d$.  Since $\Delta$ is a bounded
symmetric operator, $e^{it\Delta}$ is the unitary solution operator of the free discrete
Schr\"odinger equation
 \beq\label{eq:free-d-dim-Schroedinger}
    i\partial_t u = -\Delta u
 \eeq
on $l^2(\Z^d)$; for any $f\in l^2(\Z^d)$ the function $u(t,x)= (e^{it\Delta}f)(x)$ solves
\eqref{eq:free-d-dim-Schroedinger} and $u(0,\cdot)= f$. Note that $e^{it\Delta}$ is unitary,
in particular, $\|e^{it\Delta}f\|_2= \|f\|_2$ for all $f\in l^2(\Z^d)$.
For a diffraction profile $\wti{d}$ we set $D(t)= \int_0^t\wti{d}(\xi)\, d\xi$ and
$T_t:= e^{iD(t)\Delta}$. Thus for any
initial condition $f\in l^2(\Z^d)$ the function $u(t,\cdot)=T_tf$ solves
 \beq\label{eq:free-d-dim-DiffMS}
    i\partial_t u = -\wti{d}(t)\Delta u
 \eeq
with initial condition $u(0,\cdot)=f$. Again, $T_t$ is a unitary operator on $l^2(\Z^d)$.

For a function $f:\Z^d\to \C$, its support is given by the set
 \bdm
    \supp(f):= \{x\in \Z^d:\, f(x)\not=0\}.
 \edm
For arbitrary $A\subset \Z^d$ and $x\in \Z^d$ the distance from $x$ to $A$ is given by
 \bdm
    \dist(x,A):= \inf(|x-y|:\, y\in A)
 \edm
and for subsets $A,B\subset \Z^d$ their distance is given by
 \bdm
    \dist(A,B) := \inf(\dist(x,B):\, x\in A) = \inf (|x-y|:\, x\in A, y\in B).
 \edm
For any operator $T:l^2(\Z^d)\to l^2(\Z^d)$ we denote its kernel by
 \beq
    \la x|T|y\ra:= \la \delta_x, T\delta_y\ra
 \eeq
where $\delta_y$ is the Kronecker $\delta$-function
 \bdm
    \delta_y(x):= \left\{
                    \begin{array}{cl} 1 & \text{ if } x=y\\ 0 & \text{ if } x\not = y\end{array}
                  \right..
 \edm
In particular,
 \bdm
    T f(x)= \sum_{y\in \Z^d} \la x|T|y\ra f(y) \, .
 \edm

We use the $\lesssim$ notation in inequalities, if it is convenient not to specify any
constants in the bounds: for two real-valued functions $g,h$ defined on the same domain,
$g\lesssim h$ means that there exists a non-negative constant $C$ such that
$g(x)\le Ch(x)$ for all $x$.

The extension of the non-linear and non-local functional $\calQ$ to $l^2(\Z^d)$
is again denoted by $\calQ$,
 \beq\label{eq:calQ}
    \calQ(f_1,f_2,f_3,f_4) = \int_0^1\sum_{x\in\Z^d} \ol{(T_tf_1)(x)}(T_tf_2)(x)\ol{(T_tf_3)(x)}(T_tf_4)(x)\, dt .
 \eeq
The first problem is to show that $\calQ$ is well-defined on $l^2(\Z^d)$. Due to the next lemma this turns
out to be easier than in the continuous case.

\begin{lem} \label{lem:p-q-norm-bound}
 Let $1\le p\le q\le \infty$. Then, $l^p(\Z^d)\subset l^q(\Z^d)$ and
 \beq\label{eq:pq}
   \|f\|_q \le \|f\|_p .
 \eeq
\end{lem}
\begin{proof} Clearly $\|f\|_\infty\le \|f\|_p$ for all $f\in l^p(\Z^d)$ and all $1\le p\le \infty$.

Let $0\le s\le 1$. Then for all non-negative sequences $(a_n)_{n\in\Z^d}$,
 \beq\label{eq:s}
   \big(\sum_{n\in\Z^d} a_n\big)^s\le \sum_{n\in\Z^d} a_n^s.
 \eeq
 This follows from
 \bdm
   (a_1+a_2)^s = \frac{a_1}{(a_1+a_2)^{1-s}} + \frac{a_2}{(a_1+a_2)^{1-s}}
   \le \frac{a_1}{a_1^{1-s}} + \frac{a_2}{a_2^{1-s}}
   = a_1^s+a_2^s
 \edm
 and induction. Now let $1\le p\le q<\infty$ and $f\in l^p(\Z^d)$. Then with $s=p/q$,
 \bdm
   \|f\|_p^p = \sum_{n\in\Z^d} |f(n)|^{qs} \ge \Big( \sum_{n\in\Z^d} |f(n)|^{q} \Big)^s
   = \|f\|_q^{qs} = \| f\|_q^p ,
 \edm
 where we used \eqref{eq:s}. This finishes the proof of \eqref{eq:pq}.
\end{proof}

\begin{cor}\label{cor:boundedness} For any $f_j\in l^2(\Z^d)$, $j=1,\ldots,4$, we have
 \beq \label{eq:boundedness}
   |\calQ(f_1,f_2,f_3,f_4)|\le\int_0^1 \sum_{x\in\Z^d} \prod_{j=1}^4 |T_t f_j(x)| \, dt
   \le \prod_{j=1}^4\|f_j\|_2
 \eeq
\end{cor}
\bpf
Of course, the first inequality is just the triangle inequality. By H\"older's inequality
followed by Lemma \ref{lem:p-q-norm-bound}
 \bdm
    \sum_{x\in\Z^d} \prod_{j=1}^4 |T_t f_j(x)|
    \le \prod_{j=1}^4 \|T_t f_j\|_4
    \le \prod_{j=1}^4 \|T_t f_j\|_2
    = \prod_{j=1}^4 \| f_j\|_2
 \edm
where we also used that $T_t$ is unitary on $l^2(\Z^d)$. Thus \eqref{eq:boundedness}
follows by integrating this over $t$ on $[0,1]$.
\epf
\begin{remark}
 The bound from Corollary \ref{cor:boundedness} justifies the ad-hoc formal calculation
 \beq
    \la f, Q(f_1,f_2,f_3)\ra = \calQ(f,f_1,f_2,f_3)
 \eeq
 for all $f,f_j\in l^2(\Z^d)$ with
 \beq\label{eq:Q}
    Q(f_1,f_2,f_3):= \int_0^1 T_t^{-1}\big[ T_tf_1 \ol{T_tf_2} T_tf_3 \big]\, dt .
 \eeq
 In particular, this shows that $Q(f_1,f_2,f_3)\in l^2(\Z^d)$ whenever $f_j\in l^2(\Z^d)$, and
 $Q$ defined in \eqref{eq:Q} is a bounded three linear map from $(l^2(\Z^d))^3$ to $l^2(\Z^d)$.
 This is in contrast to the continuous case, where it is bounded only for $d=1,2$,
 \cite{100DMLee07,Stanislavova05,ZGJT01}.
\end{remark}
\begin{lem}\label{lem:free} For the kernel of the free time evolution $e^{it\Delta}$ the bound
 \beq
   \sup_{t\in[-\tau,\tau]}|\la x|e^{it\Delta}|y\ra|\le \min(1, e^{4d\tau}\frac{(4d\tau)^{|x-y|}}{|x-y|!}  )
 \eeq
 holds for all $x,y\in\Z^d$ and $ 0\le \tau<\infty$.
\end{lem}
\begin{proof}
The operator $\Delta$ is given by $\Delta f(x)= \sum_{|y-x|=1} f(y) - 2d f(x)$ and,
using, for example, the discrete Fourier transform, one sees that $0\le -\Delta\le 4d$ and $\|\Delta\|= 4d$.
Now, by symmetry of $\Delta$, $e^{it\Delta}$ is unitary, hence one always has $|\la x|e^{it\Delta}|y\ra|\le 1$
for any $x,y\in\Z^d$ and all $t$.
Since $\Delta$ is bounded, the Taylor series for the exponential yields a strongly converging series for
$e^{it\Delta}$. Thus
 \begin{align*}
   \la x| e^{{it\Delta}}|y\ra
  & = \sum_{n=0}^\infty \frac{(it)^n}{n!} \la x| \Delta^n|y\ra
    = \sum_{n=|x-y|}^\infty \frac{(it)^n}{n!} \la x| \Delta^n|y\ra
 \end{align*}
 since $\la x| \Delta^n|y\ra\not= 0$ if and only if $x$ and $y$ are connected by a path of length at most $n$,
 i.e., $|x-y|\le n$. In particular, using $\|\Delta\|=4d$,
 \begin{align*}
   |\la x| e^{{it\Delta}}|y\ra|
  & \le \sum_{n=|x-y|}^\infty \frac{|t|^n}{n!} \|\Delta\|^n
    =   \sum_{l=0}^\infty \frac{(4d|t|)^{|x-y|+l}}{(|x-y|+l)!} \\
  & \le \frac{(4d|t|)^{|x-y|}}{|x-y|!}
        \sum_{l=0}^\infty \frac{(4d|t|)^l}{l!}
    =   \frac{(4d|t|)^{|x-y|}}{|x-y|!} e^{4d|t|} .
 \end{align*}
\end{proof}

\begin{lem} \label{lem:propagation-speed} Let $s\in\N$.
 For the free time-evolution $e^{it\Delta}$ generated by $\Delta$ the bound
 \beq\label{eq:propagation-speed}
  \sum_{y:\,|x-y|\ge s} \sup_{t\in[-\tau,\tau]} |\la x| e^{it\Delta} |y\ra|^2
  \lesssim \frac{(4d\tau)^{2s}\max(s,1)^{d-1}}{(s!)^2}
 \eeq
 holds, where the implicit constant depends only on the dimension $d$ and $\tau$.
\end{lem}
\begin{proof} The number of points $y\in\Z^d$ with $|y|= \sum_{j=1}^d|y_j|= n$
can be estimated by $2^d n^{d-1}$ and with Lemma \ref{lem:free} we have
 \begin{align*}
   \sum_{y:\, |x-y|\ge s} \sup_{t\in[-\tau,\tau]} & |\la x|e^{it\Delta} |y\ra|^2
    \le e^{8d\tau} \sum_{|y|\ge s} \frac{(4d\tau)^{2|y|}}{(|y|!)^2}
    \le e^{8d\tau} 2^d\sum_{n=0}^\infty (n+s)^{d-1} \frac{(4d\tau)^{2(n+s)}}{((n+s)!)^2}\\
  & \le \frac{(4d\tau)^{2s} \max(s,1)^{d-1}}{(s!)^2}
        e^{8d\tau} 2^d\sum_{n=0}^\infty \Big(1+\frac{n}{\max(s,1)}\Big)^{d-1} \frac{(4d\tau)^{2n}}{(n!)^2} \\
  & \le \frac{(4d\tau)^{2s} s^{d-1}}{(s!)^2}
        e^{8d\tau} 2^d\sum_{n=0}^\infty (1+n)^{d-1} \frac{(4d\tau)^{2n}}{(n!)^2}
 \end{align*}
 where we also used $(n+s)!\ge n! s!$.
 Thus the inequality \eqref{eq:propagation-speed} holds with constant
 $C= e^{8d\tau} 2^d\sum_{n=0}^\infty (1+n)^{d-1} \frac{(4d\tau)^{2n}}{(n!)^2} <\infty$.
\end{proof}
\begin{remark} Estimating the number of points $y\in\Z^d$ with $|y|=n$ by
$2^d n^{d-1}$ is, of course, a gross over--counting.
A tighter estimate can be given as follows: Counting the number of different
integers $x_j\ge 0$ with $\sum_{j=1}^d x_j = n$ is equal to distributing $d-1$ separators
on $n+d-1$ places, i.e,
 \bdm
    \# \{ x\in\N_0^d :\, \sum_{j=1}^d x_j = n \} = {n+d-1 \choose d-1}
 \edm
Since we have $2$ choices for the sign, except when some coordinates are zero, we get the better bound
 \bdm
 \begin{split}
    \# \{ y\in\Z^d :\, \sum_{j=1}^d |y_j| = n \}
    &   \le 2^d {n+d-1 \choose d-1} =\frac{2^d}{(d-1)!} \prod_{j=1}^{d-1} (n+j) \\
    &   = 2^d\frac{\prod_{j=1}^{d-1} (1+j/n)}{(d-1)!} n^{d-1}
        \le \frac{2^d (1+\veps) }{(d-1)!} n^{d-1}
 \end{split}
 \edm
for some small $\veps>0$ when $n$ is large. For our purpose, the rough estimate $2^d n^{d-1}$ is good enough.
\end{remark}

In the formulation of the next lemma we need some more notation. For any $r\in \R$ let
 \beq
    \lceil r \rceil := \min (z\in\Z:\, r\le z)
 \eeq
the smallest integer greater than or equal to  $r$.
\begin{lem}[Strong Bilinear bound] \label{lem:strong-bilinear}
 There exists a constant $C$ depending only on the dimension $d$  and $\tau$ such that
 for $f_1, f_2\in l^2(\Z^d)$ and $s= \dist(\supp(f_1), \supp(f_2))$
 \beq
   \sup_{t\in[-\tau,\tau]} \| (e^{it\Delta}f_1) (e^{it\Delta} f_2) \|_2   \le
   \min(1, C\frac{\max(\lceil {s/2}\rceil,1)^{d-1} (4d\tau)^{\lceil {s/2}\rceil}}{\lceil {s/2}\rceil !})
   \|f_1\|_2 \|f_2\|_2 \, .
 \eeq
\end{lem}
\begin{proof}
First of all, note that
 \bdm
    \| (e^{it\Delta}f_1) (e^{it\Delta} f_2) \|_2^2
    = \sum_{x\in\Z^d} |e^{it\Delta}f_1(x)|^2|e^{it\Delta}f_2(x)|^2
    \le \|e^{it\Delta}f_1\|_\infty^2 \|e^{it\Delta}f_2\|_2^2 .
 \edm
Hence, using Lemma \ref{lem:p-q-norm-bound} and the unicity of $e^{it\Delta}$ on $l^2(\Z^d)$, we see
 \beq
    \| (e^{it\Delta}f_1) (e^{it\Delta} f_2) \|_2^2 \le \|e^{it\Delta}f_1\|_2^2 \|e^{it\Delta}f_2\|_2^2
    = \|f_1\|_2^2 \|f_2\|_2^2
 \eeq
uniformly in $t$. Now assume that $|t|\le \tau$. Let $I_j=\supp(f_j)$, $j=1,2$ and assume,
without loss of generality that $s=\dist(I_1,I_2)\ge 1$. Moreover we need the slightly enlarged sets
 \begin{align*}
   I_1' & := \{x:\, \dist(x,I_1)\le   \dist(x, I_2) -1 \} ,\\
   I_2' & :=  \{x:\, \dist(x,I_1)\ge \dist(x, I_2)\} .
 \end{align*}
Note that $I_j\subset I_j'$, $j=1,2$,  and $I_2'= \Z^d\setminus I_1'$. The triangle inequality gives
 \bdm
   s = \dist(I_1,I_2)\le \dist(x,I_1) + \dist(x,I_2)
   \le \left\{
        \begin{array}{cl}
            2\dist(x,I_2) -1 &\text{ if } x\in I_1' \\
            2\dist(x,I_1)    &\text{ if } x\in I_2'
        \end{array}
       \right.
 \edm
so, since the distance is always an integer, we have
 \beq\label{eq:dist}
   \min(\dist(I_1', I_2), \dist(I_2', I_1)) \ge \lceil s/2\rceil .
 \eeq
Certainly, since $I_1'\cup I_2'=\Z^d$,
 \beq\label{eq:split}
 \begin{split}
   \|(e^{it\Delta}f_1) (e^{it\Delta}f_2)\|_2^2
  &=
   \sum_{x\in I_1'} |e^{it\Delta}f_1(x)|^2 |e^{it\Delta}f_2(x)|^2
   +\sum_{x\in I_2'} |e^{it\Delta}f_1(x)|^2 |e^{it\Delta}f_2(x)|^2.
 \end{split}
 \eeq
Because of
 \bdm
   e^{it\Delta} f_j(x) = \sum_{y\in I_j} \la x|e^{it\Delta}|y\ra f_j(y) ,
 \edm
the Cauchy--Schwarz inequality implies
 \bdm
   |e^{it\Delta} f_j(x)|^2 \le \|f_j\|_2^2\sum_{y\in I_j} |\la x|e^{it\Delta}|y\ra|^2 .
 \edm
Together with \eqref{eq:dist} and the bound from Lemma \ref{lem:free}, this yields
 \beq\label{eq:split1}
  \begin{split}
    \sup_{t\in[-\tau,\tau]} &\sum_{x\in I_1'}  | e^{it\Delta}f_1(x)|^2 |e^{it\Delta}f_2(x)|^2 \\
  & \le \sup_{t\in[-\tau,\tau]}\sum_{x\in I_1'} |e^{it\Delta}f_1(x)|^2
                          \|f_2\|_2^2\sup_{x\in I_1'} \sum_{y\in I_2} |\la x|e^{it\Delta}|y\ra|^2 \\
   & \le \sup_{t\in[-\tau,\tau]}\sum_{x\in \Z^d}
            |e^{it\Delta}f_1(x)|^2 \|f_2\|_2^2
            \sup_{x\in\Z^d} \sum_{y:\,|x-y|\ge \lceil s/2\rceil}|\la x| e^{it\Delta}|y\ra|^2 \\
   & \lesssim \|f_1\|_2^2\|f_2\|_2^2 \frac{\max(\lceil {s/2}\rceil,1)^{d-1} (4d\tau)^{2\lceil s/2\rceil}}{(\lceil {s/2}\rceil !)^2} .
  \end{split}
 \eeq
An identical argument gives
 \beq\label{eq:split2}
   \sup_{t\in[-\tau,\tau]}\sum_{x\in I_2'}  | e^{it\Delta} f_1 (x)|^2 | e^{it\Delta} f_2(x)|^2
   \lesssim
   \|f_1\|_2^2\|f_2\|_2^2 \frac{\max(\lceil {s/2}\rceil,1)^{d-1} (4d\tau)^{2\lceil s/2\rceil}}{(\lceil {s/2}\rceil !)^2}.
 \eeq
The bounds \eqref{eq:split1} and \eqref{eq:split2} together with \eqref{eq:split} finish the
proof of the Lemma.
\end{proof}

\begin{lem}\label{lem:strong-multilinear} For $j\in\{1,2,3,4\}$ let $f_j\in l^2(\Z^d)$. For any choice
 $j,k\in \{1,2,3,4\}$ let $s=\dist(\supp(f_k), \supp(f_l))$. Then
 \beq
  \sup_{t\in[-\tau,\tau]}
    \sum_{x\in Z^d} \prod_{j=1}^4 \big|(e^{it\Delta} f_j)(x)\big|
  \lesssim
  \frac{\max(\lceil {s/2}\rceil,1)^{d-1} (4d\tau)^{\lceil {s/2}\rceil}}{\lceil {s/2}\rceil !}
  \prod_{j=1}^4\|f_j\|_2
 \eeq
 where the implicit constant depends only on
 the dimension $d$ and $\tau$.
\end{lem}
\begin{proof}
Follows using Cauchy-Schwarz together with Corollary \ref{cor:boundedness} and Lemma \ref{lem:strong-bilinear}.
\end{proof}

\begin{cor}\label{cor:multilinear} Let the diffraction profile obey the bound \eqref{eq:D-bounded}.
 Then with $c= 1+\ln(8d\tau)$ we have
 \beq\label{eq:multilinear1}
    |\calQ(f_1,f_2,f_3,f_4)|
    \lesssim
    e^{- s(\ln s - c )/2  + (d-1)\ln (\max(s/2,1))}  \prod_{j=1}^4 \|f_j\|_2
 \eeq
 where $s=\dist(\supp(f_k), \supp(f_l))$ for any choice $j,k\in \{1,2,3,4\}$ and the implicit
 constant depends only on the dimension $d$ and $\tau$ from \eqref{eq:D-bounded}.
\end{cor}

\begin{remark} Since for any $0<\delta<1/2$,
 \bdm
     e^{- s(\ln s - c )/2  + (d-1)\ln (\max(s/2,1))}
     \lesssim e^{- \delta s\ln s }
     = s^{-\delta s} ,
 \edm
 the bound \eqref{eq:multilinear1} implies
 \beq\label{eq:multilinear2}
   |\calQ(f_1,f_2,f_3,f_4)|
  \lesssim
   {s^{-\delta s}} \prod_{j=1}^4 \|f_j\|_2
 \eeq
 for all $0<\delta<1/2$, where the implicit constant depends only on $d$, $\delta$, and $\tau$.
\end{remark}
\begin{proof} Since \eqref{eq:D-bounded} holds, there exists $0<\tau<\infty$ such that
$|D(t)|\le \tau$ for all $0\le t\le 1$. Thus, since $T_t=e^{iD(t)\Delta}$,
 \bdm
 \begin{split}
    |\calQ(f_1,f_2,f_3,f_4)| &\le \int_0^1 \sum_{x\in\Z^d} \prod_{j=1}^4 \big|(T_t f_j)(x)\big|\, dt
    \le \sum_{x\in\Z^d} \sup_{0\le t\le 1} \prod_{j=1}^4 \big|(T_t f_j)(x)\big| \\
    &\le \sum_{x\in\Z^d} \sup_{t\in[-\tau,\tau]} \prod_{j=1}^4 \big|(e^{it\Delta} f_j)(x)\big|
 \end{split}
 \edm
and the bound from Lemma \ref{lem:strong-multilinear} implies
 \beq
    |\calQ(f_1,f_2,f_3,f_4)|
  \lesssim
   \frac{ \max(\lceil {s/2}\rceil,1)^{d-1} (4d\tau)^{\lceil {s/2}\rceil}}{\lceil {s/2}\rceil !} \prod_{j=1}^4 \|f_j\|_2  \quad.
 \eeq
An easy proof by induction shows $n! \ge e^{n\ln n -n}$. Hence, using $\lceil {s/2}\rceil\ge s/2$,
 \bdm
 \begin{split}
    \frac{\max(\lceil {s/2}\rceil,1)^{d-1} (4d\tau)^{\lceil {s/2}\rceil}}{\lceil {s/2}\rceil !}
    &\lesssim
    e^{- {s/2} \ln({s/2}) + {s/2} \ln(4d\tau) + {s/2} +(d-1)\ln (\max({s/2},1))} \\
    &=
      e^{ -s\ln(s)/2 +  c s/2 +(d-1) \ln (\max({s/2},1))}
 \end{split}
 \edm
where $c= 1+ \ln(8d\tau)$. This proves the bound \eqref{eq:multilinear1}.
\end{proof}

\section{Self-consistency bound and super-exponential decay} \label{sec:self-consistency}
As in the continuous case, see \cite{100DMLee07}, the key idea is not to focus on the
solution $f$ directly, but to study its \emph{tail distribution} defined, for $n\in\N_0$, by
 \beq\label{eq:tail}
   \alpha(n):= \Big(\sum_{|x|\ge n}|f(x)|^2\Big)^{1/2} .
 \eeq
The fundamental a-priori estimate for the tail distribution of weak solutions $f$ of
\eqref{eq:DiffMS} is given by the following
\begin{prop}[Self-consistency bound] \label{prop:self-consistency}
 Let $f$ be a weak solution of
 $\omega f= Q(f,f,f)$. Then with $c=1+\ln(8\tau)$, where $\tau$ is from the bound \eqref{eq:D-bounded}
 on the diffraction profile,
 \beq\label{eq:self-consistency-1}
    \alpha(2n)
   \lesssim
    \alpha(n)^3 + e^{- (n+1)(\ln (n+1) - c )/2}  .
 \eeq
 In particular, for any $0<\delta<1/2$ the bound
 \beq\label{eq:self-consistency-2}
   \alpha(2n)
  \lesssim
   \alpha(n)^3 + (n+1)^{-\delta (n+1)}
 \eeq
 holds. In \eqref{eq:self-consistency-1} and \eqref{eq:self-consistency-2} the
 implicit constants depend only on $\omega$, $\delta$, $\|f\|_2$, and $\tau$.
\end{prop}
\begin{proof}
 Since $f$ is a weak solution of $\omega f=Q(f,f,f)$, we have, by definition,
 \bdm
   \omega\la g,f\ra = \calQ(g,f,f,f) \quad \text{for all } g\in l^2(\Z).
 \edm
 Since
 \bdm
   \alpha(2n) = \sup_{\substack{g\in l^2(\Z), \,\|g\|_2=1,\\ \supp(g)\subset (-\infty,-2n]\cup[2n,\infty)}}|\la g,f \ra|
 \edm
 we need to estimate $\calQ(g,f,f,f)$ for $g\in l^2(\Z)$ with $\|g\|_2=1$ and
 $\supp(g)\subset (-\infty,-2n]\cup[2n,\infty)$.
 Let $I_n= \{-n+1, \ldots n-1\}$, $I_n^c$ its complement and split $f$ into its low and high parts,
 $f= f_<+ f_>$ with $f_<= f\chi_{I_n}$ and $f_>= f\chi_{I_n^c}$. Using the multi-linearity of $\calQ$,
 \beq\label{eq:separating}
   \begin{split}
     \calQ(g,f,f,f)
    &= \calQ(g, f_<, f,f) + \calQ(g,f_>,f,f) \\
    &= \calQ(g, f_<, f,f) + \calQ(g,f_>,f_<,f) + \calQ(g,f_>,f_>,f_<) + \calQ(g,f_>,f_>,f_>).
   \end{split}
 \eeq
 The last term is estimated by
 \bdm
   |\calQ(g,f_>,f_>,f_>)| \lesssim \|g\|_2 \|f_>\|_2^3  = \alpha(n)^3.
 \edm
 For the first three terms in \eqref{eq:separating} we note that each of them contains
 one $f_<$. Since $s:=\dist(\supp(g), \supp(f_<))$ is at least $n+1$, the
 enhanced multi-linear estimate \eqref{eq:multilinear1} from Corollary \ref{cor:multilinear}
 applies and gives, since $d=1$, for the first term
 \bdm
    |\calQ(g, f_<, f,f)| \lesssim
    e^{- s(\ln s - c )/2} \|g\|_2 \|f_<\|_2 \|f\|_2^2  .
 \edm
 Similar bounds hold for the second and third terms. Collecting terms and using $s\ge n+1$,
 we see
 \bdm
 \begin{split}
    \alpha(2n) &\lesssim \omega^{-1}\left(\alpha(n)^3 + e^{- (n+1)(\ln (n+1) - c )/2}
    \big( \alpha(0)^3 + \alpha(0)^2\alpha(n) +\alpha(0)\alpha(n)^2 \big)\right) \\
    &\lesssim \alpha(n)^3 +  e^{- (n+1)(\ln (n+1) - c )/2}
 \end{split}
 \edm
 since $\alpha$ is a bounded decreasing function. This proves \eqref{eq:self-consistency-1}.
 Note that the implicit constant depends only on $\omega$, $\tau$, and $\alpha(0)= \|f\|_2$.
 To prove \eqref{eq:self-consistency-2} one either argues as above, but uses \eqref{eq:multilinear2}
 instead of \eqref{eq:multilinear1}, or simply notes that
 $e^{- (n+1)(\ln (n+1) - c )/2}\lesssim (n+1)^{-\delta (n+1)}$ for any $0<\delta<1/2$.
\end{proof}

\begin{thm}[Super-exponential decay] \label{thm:super-exp-decay}
Let $\alpha$ be a decreasing non-negative function which obeys the
self-consistency bound \eqref{eq:self-consistency-1} of Proposition \ref{prop:self-consistency}
and decays to zero at infinity. Then the bound
 \bdm
    \alpha(n)\lesssim e^{-\frac{n+1}{4}\left(\ln\left(\frac{n+1}{2}\right)-c\right)}
 \edm
holds for all $n\in\N_0$. Here one can choose $c=1+\ln(8\tau)$, with $\tau$ from the bound
\eqref{eq:D-bounded} on the diffraction profile.
\end{thm}

\begin{cor}[$=$ Theorem \ref{thm:main-decay}]\label{cor:super-exp-decay}
For any weak solution of $\omega f = Q(f,f,f)$ and any $0<\mu <\tfrac{1}{4}$ the bound
 \beq
   |f(x)| \lesssim e^{-\frac{|x|+1}{4}\left(\ln\left(\frac{|x|+1}{2}\right)-c\right)}
 \eeq
holds for all $x\in\Z$.
\end{cor}
\bpf
Given Theorem \ref{thm:super-exp-decay}, this follows immediately from $|f(x)|\le \alpha(|x|)$.
\epf
It remains to prove Theorem \ref{thm:super-exp-decay}. This is done in two steps. The first is a reduction
of the full super-exponential decay to a slower but still super-exponential decay.

\begin{lem}\label{lem:reduction}
Let $\alpha$ be a non-negative decreasing function which obeys the self-consistency bound
\eqref{eq:self-consistency-1}. Then the bounds
 \beq\label{eq:reduction1}
    \alpha(n)\lesssim e^{-\frac{n+1}{4}\left(\ln\left(\frac{n+1}{2}\right)-c\right) }
 \eeq
and
 \beq\label{eq:reduction2}
    \alpha(n)\lesssim (n+1)^{-\mu_0(n+1)}
 \eeq
for some  $\mu_0>0$ are equivalent.
\end{lem}
\bpf Of course, the bound \eqref{eq:reduction1} implies \eqref{eq:reduction2} for all $0<\mu_0<1/4$.
To prove the converse we will show that if $\alpha(n)\lesssim (n+1)^{-\mu(n+1)}$ for some
$\mu>0$ and if $3\mu< 1/2$ one can boost the decay to
 \beq\label{eq:bootstrap-1-root}
    \alpha(n)\lesssim (n+1)^{-\frac{5}{4}\mu (n+1)} .
 \eeq
 Assume this for the moment and assume that \eqref{eq:reduction2} holds for some $\mu_0>0$.
 Let $l_0\in\N_0$ such that $3(5/4)^{l_0-1}\mu_0<1/2\le 3(5/4)^{l_0}\mu_0$.
 We can iterate \eqref{eq:bootstrap-1-root} exactly $l_0$ times to see
 \beq\label{eq:bootstrap-1-2}
    \alpha(n)\lesssim (n+1)^{-(5/4)^{l_0}\mu_0(n+1)} .
 \eeq
 Plugging the estimate \eqref{eq:bootstrap-1-2} into the self--consistency bound
 \eqref{eq:self-consistency-1}  yields
 \bdm
    \alpha(2n)\lesssim (n+1)^{-3(5/4)^{l_0}\mu_0(n+1)} + e^{- (n+1)(\ln (n+1) - c )/2}
    \lesssim e^{- (n+1)(\ln (n+1) - c )/2}
 \edm
 since $3(5/4)^{l_0}\mu_0\ge 1/2$, by assumption. Thus for even $n$ we have the bound
 \beq\label{eq:even}
    \alpha(n)\lesssim e^{- (\frac{n}{2}+1)(\ln (\frac{n}{2}+1) - c )/2}
    = e^{- \frac{n+2}{4}(\ln (\frac{n+2}{2}) - c )}
 \eeq
and, by monotonicity of $\alpha$, for odd $n$ the bound \eqref{eq:even} yields
 \beq\label{eq:odd}
    \alpha(n)\le \alpha(n-1)\lesssim e^{- \frac{n+1}{4}(\ln (\frac{n+1}{2}) - c )}.
 \eeq
The bounds \eqref{eq:even} and \eqref{eq:odd} together show that \eqref{eq:reduction1} holds.

 It remains to prove the boost in decay given in \eqref{eq:bootstrap-1-2}. If
 $\alpha(n)\lesssim (n+1)^{-\mu(n+1)}$ and $3\mu<1/2$, the self-consistency bound
 \eqref{eq:self-consistency-1} gives
 \bdm
    \alpha(2n)\lesssim (n+1)^{-3\mu(n+1)} + e^{- (n+1)(\ln (n+1) - c )/2}
    \lesssim (n+1)^{-3\mu(n+1)}
 \edm
 as long as $3\mu<1/2$. Thus, as before, for even $n$ one gets
 \beq\label{eq:even-2}
    \alpha(n)\lesssim \big(\frac{n+2}{2}\big)^{-\frac{3}{2}\mu(n+2)}
    \lesssim (n+2)^{-(\frac{3}{2}-\veps)(n+2)}
    \le (n+1)^{-(\frac{3}{2}-\veps)(n+1)}
 \eeq
 for any $\veps>0$. For odd $n$ the monotonicity of $\alpha$ and \eqref{eq:even-2} give
 \beq\label{eq:odd-2}
    \alpha(n)\le \alpha(n-1)\lesssim (n+1)^{-(\frac{3}{2}-\veps)(n+1)}
 \eeq
 for any $\veps>0$. The bounds \eqref{eq:even-2} and \eqref{eq:odd-2} together show
 \beq
    \alpha(n)\lesssim (n+1)^{-(\frac{3}{2}-\veps)(n+1)}
 \eeq
 for all $n\in\N_0$ and all $\veps>0$. Choosing $\veps=1/4$ yields \eqref{eq:bootstrap-1-root}.
\epf

Given Lemma \ref{lem:reduction}, in order to prove the super--exponential decay of $\alpha$
given in Theorem \ref{thm:super-exp-decay}, it is enough to show that
$\alpha(n)\lesssim (n+1)^{-\mu_0(n+1)}$ for some arbitrarily small $\mu_0>0$. This is the content of the
next proposition
\begin{prop}\label{prop:some-super-exp-decay} Assume that $\alpha$ is a non-negative decreasing function
which obeys the self-consistency bound \eqref{eq:self-consistency-1} and goes to zero at infinity. Then
there exists $\mu_0>0$ such that
 \bdm
    \alpha(n)\lesssim (n+1)^{-\mu_0(n+1)}
 \edm
\end{prop}

For the proof of Proposition \ref{prop:some-super-exp-decay} we need some more notation.
Given $n\in\N_0$ let
 \beq
   F(n):= \left\{ \begin{array}{cl}
            (n+2)^{n+2} &\text{ if } n \text{ is even } \\
            (n+1)^{n+1}         &\text{ if } n \text{ is odd }
          \end{array}\right.
 \eeq
and, for $\veps \ge 0$, its regularized version
 \beq
   F_\veps(n) := \frac{F(n)}{1+\veps F(n)} = \frac{1}{F(n)^{-1}+\veps}.
 \eeq
 Finally, for $\mu\ge 0$ let
 \beq
   F_{\mu,\veps}(n) := F_\veps(n)^\mu = (F(n)^{-1}+\veps)^{-\mu}.
 \eeq
 Furthermore let
 \beq\label{eq:norm}
    \|\alpha\|_{\mu,\veps,b} := \sup_{n\ge b} F_{\mu,\veps}(n) |\alpha(n)| .
 \eeq
Of course, the super-exponential decay given in Proposition \ref{prop:some-super-exp-decay}
is equivalent to showing
 \bdm
    \|\alpha\|_{\mu_0,0,b} <\infty \qquad\text{for some } \mu_0>0 \text{ and some } b\in\N_0 .
 \edm
Since $\|\alpha\|_{\mu,0,b}= \sup_{0<\veps\le 1} \|\alpha\|_{\mu,\veps,b}$, see
Lemma \ref{lem:F5} below, it is enough to find an $\veps$-independent bound on
$\|\alpha\|_{\mu,\veps,b}$, which is where the second self-consistency bound of
Proposition \ref{prop:self-consistency} enters.
First we gather some basic properties of $F_{\mu,\veps}$ needed in the proof of
Proposition \ref{prop:some-super-exp-decay}.
\begin{lem} \label{lem:F}
 \itemthm\label{lem:F0} For any $\veps\ge 0$ the function $n\mapsto F_\veps (n)$ is increasing
 in $n$ and, for fixed $n$, decreasing in $\veps\ge 0$. Moreover, $F_\veps(n)\le \veps^{-1}$
 for all $n$. \\[0.2em]
 \itemthm\label{lem:F1} For any $\mu,\veps\ge 0$ the function $n\mapsto F_{\mu,\veps}(n)$ is
 increasing and bounded by $\veps^{-\mu}$. Moreover,  $F_{\mu,\veps}(n)$ is decreasing in
 $\veps\ge 0$ and depends continuously on the parameters $\mu$ and $\veps$ for fixed $n\in\N_0$. \\[0.2em]
 \itemthm\label{lem:F2} For any $0\le \mu\le \delta/3$,
 the function $\N_0\ni n\to F_{\mu,0}(2n)(n+1)^{-\delta (n+1)}$ is decreasing.\\[0.2em]
 \itemthm\label{lem:F3} The bound
 \beq\label{eq:F3}
   F_{\mu,\veps} (2n) \le 4  F_{\mu,\veps} (n)^3
 \eeq
 holds for all $0\le \mu,\veps\le 1$ and $n\in\N_0$. \\[0.2em]
 \itemthm\label{lem:F4} For fixed $b\in\N_0$ and an arbitrary bounded function $\alpha$ the map
 $(\mu,\veps)\mapsto \|\alpha\|_{\mu,\veps,b}$ is continuous on $[0,1]\times (0,1]$. \\[0.2em]
 \itemthm \label{lem:F5} For fixed $0<\mu $, $b\in\N_0$, and an arbitrary bounded
 function $\alpha$,
 \bdm
    \|\alpha\|_{\mu,0,b} = \lim_{\veps\to 0}  \|\alpha\|_{\mu,\veps,b}
    =\sup_{0<\veps\le 1} \|\alpha\|_{\mu,\veps,b} .
 \edm
\end{lem}
\begin{remark} The last part of the lemma shows that for fixed $0<\mu\le 1$ and $b\in\N_0$
 the map $\veps\mapsto \|\alpha\|_{\mu,\veps,b}$ is continuous on $[0,1]$. Here we interpret
 continuity in a generalized sense:
 One certainly has $\|\alpha\|_{\mu,\veps,b}\le \|\alpha\|_\infty/\veps<\infty$ for all $\veps>0$.
 If $\|\alpha\|_{\mu,0,b}<\infty$, then
 $\lim_{\veps\to 0 }\|\alpha\|_{\mu,\veps,b} = \|\alpha\|_{\mu,0,b}$, and if
 $\|\alpha\|_{\mu,0,b} =\infty$, then $\lim_{\veps\to 0 }\|\alpha\|_{\mu,\veps,b}= \infty$.
\end{remark}
We postpone the proof of Lemma \ref{lem:F} after the proof of Proposition \ref{prop:some-super-exp-decay}.
\begin{proof}[Proof of Proposition \ref{prop:some-super-exp-decay}]
We assume that $\alpha$ decays monotonically to zero and obeys the second self-consistency bound
given in Proposition \ref{prop:self-consistency}. That is, for fixed $0<\delta<1/2$ there exists
a constant $C_0$ such that
 \beq
   \alpha(2n) \le C_0 \alpha(n)^3 +C_0 (n+1)^{-\delta (n+1)} .
 \eeq
Multiply this by $\chi_{[b,\infty)}$, the characteristic function of the set $[b,\infty)$, using
 \beq
   \chi_{[b,\infty)}(n) = \chi_{[b,\infty)}(2n) - \chi_{[b,2b)}(2n) ,
 \eeq
putting $\alpha_b = \chi_{[b,\infty)}\alpha$, and rearranging terms  yields
 \beq\label{eq:self-consistency-b}
   \alpha_b(2n) \le C_0 \alpha_b(n)^3 + C_0 (n+1)^{-\delta (n+1)} \chi_{[b,\infty)}(n)
   + \chi_{[b,2b)}(2n)\alpha(2n) .
 \eeq
Now let $\mu\le \delta/3$. Multiplying \eqref{eq:self-consistency-b} by $F_{\mu,\veps}(2n)$, using the
bound \eqref{eq:F3} from Lemma \ref{lem:F} on the first term on the right hand side of
\eqref{eq:self-consistency-b} and $F_{\mu,\veps}(2n)\le F_{\mu,0}(2n)$ from Lemma \ref{lem:F1}
on the other two, assuming $b\ge 0$, we get
 \beq
 \begin{split}\label{eq:self-consistency-F}
   F_{\mu,\veps} (2n)\alpha_b(2n)
 &\le
   C_1 \big(F_{\mu,\veps} (n)\alpha_b(n)\big)^3
   + C_0 \frac{F_{\mu, 0} (2n)}{(n+1)^{\delta (n+1)}} \chi_{[b,\infty)}(n) \\
 & \phantom{\le~}  + F_{\mu, 0} (2n) \chi_{[b,2b)}(2n)\alpha(2n) \\
 &\le
   C_1 \big(F_{\mu,\veps} (n)\alpha_b(n)\big)^3
   + C_0 \frac{F_{\mu,0} (2b)}{(b+1)^{\delta (b+1)}}
   + F_{\mu,0}(2b) \alpha(b) \\
 &\le
    C_1 \|\alpha\|_{\mu,\veps,b}^3
    +  F_{\mu,0} (2b) \left( \frac{C_0}{(b+1)^{\delta (b+1)}} + \alpha(b) \right) .
 \end{split}
 \eeq
For the second inequality we used that $\alpha$ and, by Lemma \ref{lem:F2}, $F_{\mu,0}(2n)(n+1)^{-\delta (n+1)}$
are decreasing and that $F_{\mu,0}(2n)$ is increasing in $n$. In the third inequality, we used
the obvious bound
 \bdm
   F_{\mu,\veps}(n)\alpha_b(n) \le \|\alpha\|_{\mu,\veps,b} \quad \text{for all } n\in\N_0 .
 \edm
by the definition \eqref{eq:norm} for $\|\alpha\|_{\mu,\veps,b}$.

The punchline is that, since $F_{\mu,\veps}(2n)= F_{\mu,\veps}(2n+1)$ by construction of
$F_{\mu,\veps}$, the monotonicity of $\alpha$ gives
 \beq
   \sup_{n\ge b} F_{\mu,\veps}(n)\alpha(n) = \sup_{n\in\N_0} F_{\mu,\veps}(2n)\alpha_b(2n)
 \eeq
whenever $b$ is an even natural number. So, since \eqref{eq:self-consistency-F} holds for any
$n\in\N_0$, taking the supremum over $n$ in \eqref{eq:self-consistency-F} yields
 \beq\label{eq:self-consistency-closed}
   \|\alpha\|_{\mu,\veps,b}
   \le
   C_1 \|\alpha\big\|_{\mu,\veps,b}^3
   +  F_{\mu,0} (2b) \left( \frac{C_0}{(b+1)^{\delta (b+1)}} + \alpha(b) \right)
 \eeq
which is a \emph{closed} inequality for $\|\alpha\|_{\mu,\veps,b}$ and holds
as long as $b$ is even and $0< \mu\le \delta/3$ and $0<\veps\le 1$ or
$\mu=0$ and $0\le \veps\le 1$.

Equivalently, with $G(\nu):= \nu-C_1\nu^3$ defined for $\nu\ge 0$, we arrived at
the a-priori bound
 \beq\label{eq:self-consistency-G}
   G(\|\alpha\|_{\mu,\veps,b})
   \le F_{\mu,0} (2b) \big( C_0 (b+1)^{-\delta (b+1)} + \alpha(b) \big) .
 \eeq
A simple exercise shows that the maximum of $G$ is attained at $\nu_\text{max}= (3C_1)^{-1/2}$ and is
given by $G_{\text{max}}= G(\nu_\text{max}) = 2/(3\sqrt{3C_1})$.
Let $0<G_0<G_\text{max}$ and $0<\nu_0<\nu_\text{max}<\nu_1$ with
$G(\nu_0) = G(\nu_1) = G_0$. Since $G(\nu)<\nu$ for $\nu>0$, we have $G(\nu_0)< \nu_0$. Moreover,
 \beq\label{eq:connected-components}
   G^{-1}((-\infty,G_0]) = [0,\nu_0]\cup[\nu_1,\infty).
 \eeq
This situation is visualized in Figure \ref{fig:x-xcubed}.
\begin{figure}[htb]
\psfrag{a1}{$G_\text{max}$}
\psfrag{a0}{$G_0$}
\psfrag{bm}{$\nu_\text{max}$}
\psfrag{b0}{$\nu_0$}
\psfrag{b1}{$\nu_1$}
\psfrag{0}{$0$}
\psfrag{text}{$[0,\nu_0]\cup[\nu_1,\infty)= G^{-1}\big((-\infty,G_0]\big)$}
\includegraphics[height=6cm,width=12cm]{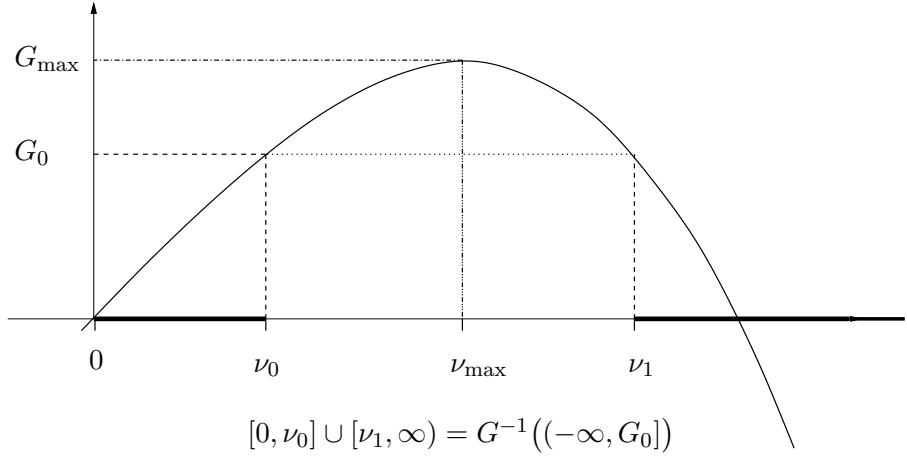}
\caption{Graph of $G(\nu)=\nu-C_1\nu^3$ and the
trapping region $G^{-1}\big((-\infty,G_0]\big)$.}
  \label{fig:x-xcubed}
\end{figure}
Now we finish the proof of the decay estimate: we need to show that $\|\alpha\|_{\mu,0,b}$
is finite for some $\mu>0$ and $b\in\N_0$.
Note that by Lemma \ref{lem:F4} the map $(\mu,\veps)\mapsto \|\alpha\|_{\mu,\veps,b}$ is continuous in
$(\mu,\veps)\in [0,1]\times(0,1]$, and, by Lemma \ref{lem:F5}, for fixed $0<\mu\le 1$
$\|\alpha\|_{\mu,0,b}= \lim_{\veps\to 0}\|\alpha\|_{\mu,\veps,b}$.
So it will be enough to find, for some $\mu>0$ and $b\in\N_0$, a uniform in $0<\veps\le 1$  estimate
for $\|\alpha\|_{\mu,\veps,b}$. This is where the bound \eqref{eq:self-consistency-G}
will enter.

\noindent
Step 1: Choose an even $b$ such that $C_0(b+1)^{-\delta (b+1)} + \alpha(b)< G_0<G_\text{max}$.
This is possible since $\alpha$ goes to zero at infinity. Since $\alpha$ is monotone decreasing,
it also guarantees $\|\alpha\|_{0,1,b}= \sup_{n\in\N_0}\alpha_b(n)= \alpha(b)< G_0 \le \nu_0$.

\noindent
Step 2: For the $b$ fixed in Step 1, let $0<\mu_0\le \delta/3$ such that
$ F_{\mu_0, 0}(2b) (C_0 (b+1)^{-\delta (b+1)} + \alpha(b))\le G_0<G_\text{max}$ and
$\|\alpha\|_{\mu_0,1,b}< \nu_0$.
This is possible since $F_{0,0}(2b)=1$ and $F_{\mu,0}(2b)$ and $ \|\alpha\|_{\mu,1,b}$ are
continuous in $0\le \mu\le \delta/3$.

Putting things together, \eqref{eq:self-consistency-G} gives
 \beq\label{eq:trapped}
    G(\|\alpha\|_{\mu_0,\veps,b}) \le G_0 \quad\text{for all } 0<\veps\le 1 .
 \eeq
Since $G$ is continuous and $\|\alpha\|_{\mu_0,\veps,b}$ depends continuously on
$\veps>0$, the bound \eqref{eq:trapped} shows that $\|\alpha\|_{\mu_0,\veps,b}$ is
trapped in the same connected component of $G^{-1}((-\infty,G_0])$ as
$\|\alpha\|_{\mu_0,1,b}$ for all $0<\veps\le 1$.
Thus, using $0\le \|\alpha\|_{\mu_0,1,b}< \nu_0$ and \eqref{eq:connected-components}, we must have
 \beq\label{eq:crucial-bound}
   \|\alpha\|_{\mu_0,\veps,b}\le \nu_0 \quad\text{for all } 0<\veps\le 1 .
 \eeq
Together with $\|\alpha\|_{\mu_0,0,b}= \lim_{\veps\to 0}\|\alpha\|_{\mu_0,\veps,b}$, the bound
\eqref{eq:crucial-bound} shows $ \|\alpha\|_{\mu_0,0,b} \le \nu_0 <\infty $
which proves the estimate
 \bdm
   \alpha(n)\le \nu_0 F(n)^{-\mu_0}
 \edm
 for all large $n$. This finished our proof of Proposition \ref{prop:some-super-exp-decay}.
\end{proof}

It remains to prove the properties of $F_{\mu,\veps}$ given in Lemma \ref{lem:F}.
\begin{proof}[Proof of Lemma \ref{lem:F}] The function $F$ is clearly increasing in $n$
and since $s\mapsto s/(1+\veps s)$ is increasing in $s\ge 0$ for fixed $\veps\ge 0$, we see
that $F_{\veps}$ and hence also $F_{\mu,\veps}$ is increasing in $n$. The other claims in
part (i) and (ii) of Lemma \ref{lem:F} are obvious. \\[0.2em]
(iii): With $\lambda= \tfrac{\delta}{2\mu} -1\ge 1/2$,  we have
 \bdm
 \begin{split}
    F_{\mu,0}(2n)(n+1)^{-\delta(n+1)}
    &= (2(n+1))^{2\mu(n+1)} (n+1)^{-\delta(n+1)}
    = \left[ 2^{n+1} (n+1)^{-\lambda(n+1)} \right]^{2\mu} .
 \end{split}
 \edm
Hence
 \bdm
    \frac{F_{\mu,0}(2(n+1))(n+2)^{-\delta(n+2)}}{F_{\mu,0}(2n)(n+1)^{-\delta(n+1)}}
        =
    \left[ \frac{2}{(n+2)^{\lambda}}\left(\Big(1+\frac{1}{n+1}\Big)^{(n+1)}\right)^{-\lambda} \right]^{2\mu} .
 \edm
Since the sequence $(1+\frac{1}{n+1})^{n+1}$ is increasing, one has $(1+\frac{1}{n+1})^{n+1}\ge 2$
and
 \bdm
    \frac{F_{\mu,0}(2(n+1))(n+2)^{-\delta(n+2)}}{F_{\mu,0}(2n)(n+1)^{-\delta(n+1)}}
        \le
    \left[ \frac{2}{(n+2)^{\lambda}}2^{-\lambda} \right]^{2\mu}
        \le
     2^{(1-2\lambda)2\mu} \le 1
 \edm
so the function $F_{\mu,0}(2n)(n+1)^{-\delta(n+1)}$ is decreasing on $\N_0$. \\[0.2em]
(iv): Put
 \beq\label{eq:f-mu-veps}
    f(n,\veps):= \frac{F_\veps(2n)}{F_\veps(n)^3} = \frac{(F(n)^{-1}+\veps)^3}{F(2n)^{-1}+\veps}.
 \eeq
 We claim that
 \beq\label{eq:constant}
    \sup_{n\in\N_0,\, {0\le\veps\le 1} } f(n,\veps)  = 4,
 \eeq
which obviously yields $F_\veps(2n)\le 4 F_\veps(n)^3$ and
 \bdm
    F_{\mu,\veps}(2n)\le 4^\mu F_{\mu,\veps}(n)^3 \le 4 F_{\mu,\veps}(n)^3
    \quad \text{ for all } 0\le \mu\le 1.
 \edm
So in order to prove \eqref{eq:F3} it is enough to show that \eqref{eq:constant} holds.
The partial derivative of $f$ with respect to $\veps$ is given by
 \bdm
    \frac{\partial f}{\partial\veps}
    =
    \frac{(F(n)^{-1}+\veps)^2}{(F(2n)^{-1}+\veps)^2}
    \big[ 3F(2n)^{-1} - F(n)^{-1}+2\veps \big].
 \edm
 In the case $3F(2n)^{-1} - F(n)^{-1} \ge 0$ one has $\tfrac{\partial f}{\partial\veps}\ge 0$
 for all $\veps \ge 0$ and the case $3F(2n)^{-1} - F(n)^{-1} < 0$ one has $\tfrac{\partial f}{\partial\veps}<0$
 as long as $0<\veps< (F(n)^{-1}-3F(2n)^{-1})/2$ and $\tfrac{\partial f}{\partial\veps}>0$ for
 $\veps > (F(n)^{-1}-3F(2n)^{-1})/2$. Altogether, as a  function of $\veps$, $f(n,\veps)$ is either increasing
 on $[0,\infty)$ or it has a single minimum and no maximum in $(0,\infty)$. Hence, for fixed $n\in\N_0$,
 its maximum in $\veps\in [0,1]$ is attained at the boundary,
 \beq\label{eq:supremum-0}
  \sup_{n\in\N_0,\,0\le \veps\le 1} f(n,\veps) = \max (\sup_{n\in\N_0}f(n,0), \sup_{n\in\N_0}f(n,1)) .
 \eeq
 Since $F_1(n)= F(n)/(1+F(n))=(1+F(n)^{-1})^{-1}$ and $F(n)\ge 4$ for all $n\in\N_0$,
 \beq\label{eq:supremum-1}
    f(n,1) =\frac{F_1(2n)}{F_1(n)^3}
    = \frac{F(2n)}{1+F(2n)} \big(1+F(n)^{-1}\big)^3
    \le \Big(\frac{5}{4}\Big)^3 <2 \,
 \eeq
and using $(n+1)^{n+1}\le F(n)\le (n+2)^{n+2}$ one sees
 \beq\label{eq:supremum-2}
    f(n,0) =\frac{F(2n)}{F(n)^3} \le \frac{\big(2(n+1)\big)^{2(n+1)}}{(n+1)^{3(n+1)}}
    = \left(\frac{4}{n+1}\right)^{n+1}
    \le 4
 \eeq
for all $n\in\N_0$. Putting \eqref{eq:supremum-0}, \eqref{eq:supremum-1}, and \eqref{eq:supremum-2}
together and noticing $f(1,0)=4$ yields \eqref{eq:constant}. \\[0.2em]
(v): Continuity of $\|\alpha\|_{\mu,\veps,b}$ in $(\mu,\veps)\in [0,1]\times(0,1]$:
 First note that the triangle inequality implies
 \begin{align}
    \big| \|\alpha\|_{\mu_1,\veps_1,b} -\|\alpha\|_{\mu_2,\veps_2,b} \big|
    &\le \sup_{n\ge b} \big| (F_{\mu_1,\veps_1}(n)- F_{\mu_2,\veps_2}(n)) \alpha(n)\big| \nonumber \\
    &\le \|\alpha\|_\infty
        \sup_{n\in\N_0} \Big| \big(F(n)^{-1}+\veps_1\big)^{-\mu_1} - \big(F(n)^{-1}+\veps_2\big)^{-\mu_2} \Big| \nonumber\\
    &\le \|\alpha\|_\infty
        \sup_{0\le x\le 1/4} \Big| \big(x+\veps_1\big)^{-\mu_1} - \big(x+\veps_2\big)^{-\mu_2} \Big|
        \label{eq:bound-in-mu-veps}
 \end{align}
 since $ 4\le F(n)$ for all $n\in\N_0$. Let $h(x,\mu,\veps):= (x+\veps)^{-\mu}$. For any $0<\veps'<1$,
 $h$ is continuous on the compact set $[0,1/4]\times[0,1]\times[\veps',1]$ and hence uniformly continuous.
 Thus for any $\eta>0$ there exists $\delta>$ such that for
 $(x_j,\mu_j,\veps_j)\in [0,1]\times[0,1]\times[\veps',1]$, $j=1,2$ with
 $|x_1-x_2|, |\mu_1-\mu_2|,|\veps_1-\veps_2|<\delta$ we have $|h(x_1,\mu_1,\veps_1)-h(x_2,\mu_2,\veps_2)|<\eta$.
 In particular,
 \bdm
    \sup_{0\le x\le 1/4} |h(x,\mu_1,\veps_1)-h(x,\mu_2,\veps_2)|<\eta
 \edm
 which, together with \eqref{eq:bound-in-mu-veps}, shows that
 $(\mu,\veps)\mapsto \|\alpha\|_{\mu,\veps,b}$ is uniformly continuous on any compact subset
 of $[0,1]\times (0,1]$, hence continuous on $[0,1]\times (0,1]$. \\[0.2em]
(vi): Fix $\mu>0$. Recall that $\|\alpha\|_{\mu,\veps,b}$ is decreasing in $\veps$. Thus
 \bdm
  \begin{split}
    \lim_{\veps\to 0}\|\alpha\|_{\mu, \veps ,b}
   &= \sup_{0<\veps\le 1}  \|\alpha\|_{\mu, \veps ,b}
      = \sup_{0<\veps\le 1} \sup_{n\ge b} F_{\mu,\veps}(n) |\alpha(n)| \\
   &= \sup_{n\ge b} \sup_{0<\veps\le 1} F_{\mu,\veps}(n) |\alpha(n)|
      = \sup_{n\ge b} F_{\mu,0}(n) |\alpha(n)|
    = \|\alpha\|_{\mu,0,b}
  \end{split}
 \edm
 which finishes the proof of Lemma \ref{lem:F}.
\end{proof}

\section{Existence of Diffraction managed solitons for zero average diffraction} \label{sec:existence}
Here we want to give a simple proof of existence of diffraction managed solitons, i.e., weak solutions of
\eqref{eq:DiffMS}, via the direct
method from the calculus of variations. This hinges on the fact that the diffraction management
equation is the Euler--Lagrange equation for the functional
 \beq\label{eq:varphi}
   \varphi(f):= \calQ(f,f,f,f)
 \eeq
 on $l^2(\Z)$. The corresponding constraint maximization problem
 is given by
 \beq\label{eq:variational}
   P_\lambda= \sup\big( \varphi(f):\, f\in l^2(\Z), \|f\|_2^2=\lambda \big)
 \eeq
 where $\lambda>0$.
 Up to some minor technical details it is clear that any maximizer $f$ of the variational problem
 \eqref{eq:variational}, that is, any $f\in l^2(\Z)$ with $\|f\|_2^2=\lambda$ and
 \beq\label{eq:maximizer}
    \calQ(f,f,f,f) = P_\lambda
 \eeq
 is by the Lagrange multiplier theorem a weak solution of the
 diffraction management equation \eqref{eq:DiffMS}.

The usual way to show existence of a maximizer is to identify it as the limit of a suitable
maximizing sequence, i.e., a sequence $(f_n)_n$ with $\|f_n\|_2^2=\lambda$ and
$\lim_{n\to\infty }\varphi(f_n)= P_\lambda$. Such a sequence always exists, the problem,
due to the translation invariance of the time evolution $T_t= e^{iD(t)\Delta}$, is that
the functional $\varphi$ is invariant under translation; if $f_n$ is a maximizing
sequence for \eqref{eq:variational} and $\xi_n$ any sequence in $\Z$, then the shifted sequence
 \bdm
   f_{n,\xi_n}(x):= f_n(x-\xi_n)
 \edm
is also a maximizing sequence. That is, the problem \eqref{eq:variational} is invariant under
shifts, yielding a loss of compactness since maximizing sequences can very easily converge
weakly to zero. The usual way to overcome this is the use of
Lions' concentration compactness method \cite{Lions84} which, for positive average dispersion
has been used in \cite{Moeser05,Panayotaros05}. For vanishing average diffraction, and under
much more restrictive conditions on the diffraction profile than \eqref{eq:D-bounded}, the existence
of a maximizer of \eqref{eq:variational} has been shown in \cite{Stanislavova07} with the help of
Ekeland's variational principle, \cite{Ekeland72,Ekeland73}, see also \cite{Jabri}.
We will give a different approach, which avoids the use of Lions' concentration compactness method
or Ekeland's variational principle by using the translation invariance of the problem to show that
for any maximizing sequence $f_n$ there exists a sequence of shifts $\xi_n$ such that the
shifted sequence $f_{n,\xi_n}$ is tight in the sense of \eqref{eq:tight} below.
Then, since $f_{n,\xi_n}$ is bounded in $l^2(\Z)$, it has a weakly
convergent and hence, by the compactness Lemma \ref{lem:compactness-criterion} below, also a
strongly convergent subsequence. The limit of this subsequence is then a natural candidate for
the maximizer of \eqref{eq:variational}, see Theorem \ref{thm:existence}.

In the following, we will always assume that the diffraction profile obeys the
bound \eqref{eq:D-bounded}.
Our main tools are the multi-linear bound from Corollary \ref{cor:multilinear}
and the following simple compactness result.

\begin{lem}\label{lem:compactness-criterion} Let $1\le p<\infty$.
A sequence $(f_n)_{n\in\N}\subset l^p(\Z^d)$ is strongly converging to $f$ in $l^p(\Z^d)$
if and only if it is weakly convergent to $f$ and the sequence is tight, i.e.,
 \beq\label{eq:tight}
   \lim_{L\to\infty} \limsup_{n\to\infty} \sum_{|x|>L} |f_n(x)|^p = 0.
 \eeq
\end{lem}
\begin{proof} For $L\ge 1$, let $K_L$ be the operator of multiplication with the
characteristic function of $[-L,L]^d\cap\Z^d$, that is,
 \bdm
   K_L f(x)= \left\{
                \begin{array}{cl}
                    f(x) & \text{ for } |x|\le L \\
                    0    & \text{ for } |x|> L
                \end{array}
             \right.
 \edm
and $\ol{K}_L:= \id-K_L$. Note that both $K_L$ is an operator with finite dimensional range.
In particular, for any $1\le L<\infty$, $K_L:l^p(\Z^d)\to l^p(\Z^d)$ is a compact operator.
Furthermore, both $K_L$ and $\ol{K}_L$ are bounded operators with operator norm
one since
 \bdm
    \|f\|_p^p = \|K_L f\|_p^p + \|\ol{K}_L f\|_p^p \ge \big[\max\big(\|K_L f\|_p, \|\ol{K}_L f\|_p  \big)\big]^p
 \edm
for all $f\in l^p(\Z^d)$ and $K_L$, respectively $\ol{K}_L$, acts as the identity on its range.
Moreover, the
tightness-condition \eqref{eq:tight} is equivalent to
 \beq\label{eq:tight-2}
   \lim_{L\to\infty}\limsup_{n\to\infty}\|\ol{K}_L f_n\|_p = 0 .
 \eeq\\[0.2em]
Now assume that  $f_n$ converges strongly to $f$ in $l^p(\Z^d)$. Then it clearly converges also weakly to
$f$ and
 \bdm
   \|\ol{K}_L f_n\|_p \le \|\ol{K}_L f\|_p + \|\ol{K}_L (f_n-f)\|_p \le \|\ol{K}_L f\|_p + \|f_n-f\|_p  .
 \edm
Thus, by strong convergence of $f_n$ to $f$,
 \bdm
   \limsup_{n\to\infty} \|\ol{K}_L f_n\|_p \le  \|\ol{K}_L f\|_p \,\, .
 \edm
Since $f\in l^p(\Z^d)$ and $p<\infty$, we have $\lim_{L\to\infty} \|\ol{K}_L f\|_p = 0$, so the sequence
$f_n$ is tight. \\[0.4em]
For the converse assume that $f_n$ converges to $f$ weakly in $l^p(\Z^d)$ and is tight, that is,
\eqref{eq:tight-2} holds. Then
 \bdm
   \|f-f_n\|_p \le \|K_L(f-f_n)\|_p  + \|\ol{K}_L(f-f_n)\|_p
   \le \|K_L(f-f_n)\|_p  + \|\ol{K}_L f\|_p   + \|\ol{K}_L f_n\|_p \, .
 \edm
 Since $K_L$ is a compact operator on $L^p$, it maps weakly convergent sequences
 into strongly converging sequences. Hence
 \bdm
   \limsup_{n\to\infty}\|f-f_n\|_p
   \le \|\ol{K}_L f\|_p   +  \limsup_{n\to\infty} \|\ol{K}_L f_n\|_p
 \edm
 Now using $f\in l^p(\Z^d)$ and the condition \eqref{eq:tight-2}, take the limit $L\to\infty$ in this
 inequality to get
 \bdm
   \limsup_{n\to\infty}\|f-f_n\|_p \le 0 .
 \edm
 Thus $f_n$ converges to $f$ strongly in $l^p(\Z^d)$.
\end{proof}

\begin{remark} The proof above breaks down for $p=\infty$, since for an arbitrary $f\in l^\infty(\Z^d)$,
one does not, in general, have
 \bdm
    \lim_{L\to\infty} \|\ol{K}_L f\|_\infty = 0.
 \edm
However, for the closed subspace $l^\infty_0(\Z^d)\subset l^\infty(Z^d)$ consisting of bounded sequences
indexed by $\Z^d$ vanishing at infinity, $\lim_{x\to\infty} f(x)=0$ for any $f\in l^\infty(\Z^d)$, the above
proof immediately generalizes and yields the analogous compactness statement: $(f_n)_{n\in\N}$ converges strongly
in $l^\infty_0(\Z^d)$ if and only if it converges weakly and
 \bdm
    \lim_{L\to\infty} \limsup_{n\to\infty} \sup_{|x|>L} |f_n(x)| = 0.
 \edm
\end{remark}
In the next Lemma we gather some simple bounds for the non-linear functional
$\varphi$ and the associated constraint maximization problem \eqref{eq:variational}.
\begin{lem} \label{lem:scaling}
For any $\lambda>0$ one has $0< P_\lambda\le 1$ and  the scaling property
$P_\lambda= P_1\lambda^2$ holds. In particular, for any $f\in l^2(\Z)$ the bound
 \bdm
    \varphi(f) \le P_1 \|f\|_2^4
 \edm
holds.
\end{lem}
\bpf Since $\varphi(f)=\calQ(f,f,f,f)= \int_0^1 \|T_t f\|_4^4\, dt$ we obviously have
$P_\lambda\ge \varphi(f)\ge 0$. With Corollary \ref{cor:boundedness}, we see
$P_\lambda\le 1$. By scaling, replacing $f$ with $\wti{f}= f/\sqrt{\lambda}$, one gets
 \bdm
    \calQ(f,f,f,f) = \lambda^2 \calQ(\wti{f},\wti{f},\wti{f},\wti{f})
 \edm
and taking the supremum over $f$ with $\|f\|_2^2=\lambda$ we see $P_\lambda= P_1\lambda^2$.
If $P_1=0$ then  $ 0= \calQ(f,f,f,f) = \int_0^1 \|T_tf\|_4^4\, dt $
for all $f\in l^2(\Z)$ with $\|f\|_2=1$.
Thus, for almost every $t\in[0,1]$ one would have $T_tf(x)=0$ for all $x\in\Z$. Hence $f=0$ in
contradiction to $\|f\|_2=1$. Thus $P_1>0$.
\epf

The next lemma is key in our proof that maximizing sequences have strongly convergent subsequences
modulo translations. Its proof is inspired by the proof of Lemma 2.3 in \cite{KMZ05}.

\begin{lem}\label{lem:epsilon-fat-tail} Assume that the diffraction profile obeys the bound
\eqref{eq:D-bounded}. Then there is a constant $C$ such that if $f\in l^2(\Z)$,
$\veps^2<\|f\|_2^2$ and $a,b\in\Z$, $a\le b$ with
 \beq
 \begin{split} \label{eq:epsilon-fat-tail}
   \sum_{x < a} |f(x)|^2 \ge \frac{\veps^2}{2} \text{ and }
   \sum_{x > b} |f(x)|^2 \ge \frac{\veps^2}{2} ,
 \end{split}
 \eeq
 then
 \beq\label{eq:epsilon-fat-tail-bound}
   \varphi(f) \le P_1(\|f\|_2^4 -\veps^4/2) +\frac{C \|f\|_2^4}{ \big((b-a+1)^{1/2}-1\big)^{1/2}}
 \eeq
\end{lem}
\bpf The number of points in $[a,b]\cap\Z$ is given by $b-a+1$. Choose $l\in\N$ such that
 \beq\label{eq:l}
   l\le (b-a+1)^{1/2} < l+1
 \eeq
and let $I$ be a subinterval of $\{a,a+1,\ldots,b\}$ consisting of $l^2$ consecutive points.
By pigeonholing, there must be a subset $I_0= \{a',\ldots,b'\}\subset I$ consisting of
$l$ consecutive points with
 \beq \label{eq:small-middle-part}
   \sum_{x\in I_0} |f(x)|^2 \le \frac{\|f\|_2^2}{l} .
 \eeq
We split $f$ into three parts, $f_{-1}:= f\vert_{(-\infty,a')}$, $f_{1}:= f\vert_{(b',\infty)}$, and
$f_0:= f\vert_{I_0}$. Then $f=f_{-1}+f_0+f_{1}$,
$\|f\|_2^2 = \|f_{-1}\|_2^2 + \|f_{0}\|_2^2 + \|f_{1}\|_2^2$, and \eqref{eq:epsilon-fat-tail} and
\eqref{eq:small-middle-part} give
 \beq \label{eq:fat-small}
    \|f_{-1}\|_2^2\ge \frac{\veps^2}{2}, \quad \|f_1\|_2^2\ge \frac{\veps^2}{2}, \quad
    \text{and } \|f_0\|_2 \le \frac{\|f\|_2}{\sqrt{l}} .
 \eeq

Using the multi-linearity of $\calQ$,
 \beq\label{eq:refined-boundedness1}
 \begin{split}
   \calQ(f,f,f,f)
  &= \sum_{\substack{j_l\in\{-1,0,1\} \\ l=1,\ldots,4}} \calQ(f_{j_1},f_{j_2},f_{j_3},f_{j_4}) \\
  &= \sum_{\substack{ j_l\in\{-1,1\} \\ l=1,\ldots,4}} \calQ(f_{j_1},f_{j_2},f_{j_3},f_{j_4})
        + \sum_{\substack{ j_l\in\{-1,0,1\} \\  l=1,\ldots,4 \\ \text{some } j_l=0} }
          \calQ(f_{j_1},f_{j_2},f_{j_3},f_{j_4}) \\
  &= \calQ(f_{-1},f_{-1},f_{-1},f_{-1}) + \calQ(f_{1},f_{1},f_{1},f_{1}) \\
  &     + \sum_{\substack{ j_l\in\{-1,1\} \\ \exists k,l: j_k=-1, j_l=1 }}\calQ(f_{j_1},f_{j_2},f_{j_3},f_{j_4})
        + \sum_{\substack{ j_l\in\{-1,0,1\} \\ l=1,\ldots,4 \\ \text{some } j_l=0}}
          \calQ(f_{j_1},f_{j_2},f_{j_3},f_{j_4}).
 \end{split}
 \eeq
 Lemma \ref{lem:scaling} shows
 \beq\label{eq:first-two}
   \calQ(f_{-1},f_{-1},f_{-1},f_{-1}) + \calQ(f_{1},f_{1},f_{1},f_{1})
   = \varphi(f_{-1}) + \varphi(f_1)
   \le P_1 \big( \|f_{-1}\|_2^4 + \|f_{1}\|_2^4 \big)
 \eeq
 and, utilizing that the supports of $f_{-1}$ and $f_1$ have distance at least $l+1\ge 2$,
 the enhanced multi-linear bound  \eqref{eq:multilinear2} with the choice $\delta=1/4$
 gives
 \beq\label{eq:third}
    \begin{split}
        \sum_{\substack{ j_l\in\{-1,1\} \\ \exists k,l: j_k=-1, j_l=1 }}
        \big|\calQ(f_{j_1},f_{j_2},f_{j_3},f_{j_4})\big|
    &\lesssim \frac{1}{(l+1)^{\delta(l+1)}} \|f\|_2^4
        \le \frac{\|f\|_2^4}{\sqrt{l}}  \, .
    \end{split}
 \eeq
 Also,
 \beq\label{eq:fourth}
    \sum_{\substack{j_l\in\{-1,0,1\} \\  l=1,\ldots,4 \\ \text{some } j_l=0}}
        \big|\calQ(f_{j_1},f_{j_2},f_{j_3},f_{j_4})\big|
    \le \sum_{\substack{j_l\in\{-1,0,1\} \\  l=1,\ldots,4 \\ \text{some } j_l=0}}
    \prod_{l=1}^4 \|f_{j_l}\|_2
    \le \frac{\|f\|_2^4}{\sqrt{l}}
 \eeq
 by the a-priori-bound on $\calQ$ given in Corollary \ref{cor:boundedness}
 and \eqref{eq:fat-small} since each term in the above sum
 contains at least one factor $\|f_0\|_2$.
 Thus \eqref{eq:refined-boundedness1} together with \eqref{eq:first-two}, \eqref{eq:third}, and
 \eqref{eq:fourth} gives
 \beq\label{eq:refined-boundedness2}
   \varphi(f)= \calQ(f,f,f,f)
   \le  P_1\big( \|f_{-1}\|_2^4 + \|f_{1}\|_2^4 \big)
        + \frac{C}{\sqrt{l}}\|f\|_2^4
 \eeq
 for some constant $C$. Since $\|f_{-1}\|_2^2,\|f_{1}\|_2^2\ge \veps^2/2$ and
 $\|f_{-1}\|_2^2+\|f_{1}\|_2^2\le \|f\|_2^2$, we have
 \bdm
   \|f_{-1}\|_2^4+\|f_{1}\|_2^4 \le \|f\|_2^4 - 2\|f_{-1}\|_2^2 \|f_1\|_2^2 \le \|f\|_2^4 - \veps^4/2.
 \edm
 Together with \eqref{eq:l} the bound
 \eqref{eq:refined-boundedness2} yields \eqref{eq:epsilon-fat-tail-bound}.
\epf

\begin{lem}\label{lem:maximizing-seq-tight} Assume that the diffraction profile obeys
the bound \eqref{eq:D-bounded} and let $(f_n)_{n\in\N}$ be a maximizing sequence for
the constraint maximization problem \eqref{eq:variational}.
Then there exists a sequence  $(\xi_n)_{n\in\N}$
and, for each $0<\veps<\sqrt{\lambda}$, there exists $R_\veps$, which is independent of $n$,
such that
 \beq\label{eq:maximing-seq-tight-modulo-shift}
   \limsup_{n\to\infty} \sum_{|x -\xi_n|>R_\veps} |f_n(x)|^2 \le \veps^2 .
 \eeq
\end{lem}
\bpf
Given $0<\veps<\sqrt{\lambda}$ as in the Lemma, we will show that there are corresponding
intervals $I_{n,\veps}= \{c_{n,\veps},\ldots, d_{n,\veps}\}$ which are nested,
 \beq\label{eq:nested}
    I_{n,\veps}\subset I_{n,\veps'} \quad\text{ for all } 0<\veps'\le\veps,
 \eeq
have bounded length,
 \beq\label{eq:finite-length}
    \sup_{n\in\N} (d_{n,\veps} -c_{n,\veps}) <\infty,
 \eeq
and contain most of the $l^2$ norm of $f_n$,
 \beq\label{eq:I-n-veps}
    \limsup_{n\to\infty} \sum_{x\not\in I_{n,\veps}} |f_n(x)|^2 \le \veps^2 .
 \eeq
Given \eqref{eq:nested}, \eqref{eq:finite-length}, and \eqref{eq:I-n-veps},
$\xi_n$ and $R_\veps$ can be constructed as follows: Fix some $0<\veps_0<\sqrt{\lambda}$.
For each $n\in\N$ define $\xi_n$ by simply choosing some point from
the set $I_{n,\veps_0}$ and define $R_\veps$ by
 \bdm
    R_\veps := \sup_{n\in\N} (d_{n,\veps} -c_{n,\veps})
 \edm
 for $\veps\le \veps_0$ and $R_\veps= R_{\veps_0}$ for $\veps_0<\veps<\sqrt{\lambda}$. The bound
 \eqref{eq:finite-length} guarantees that $R_\veps$ is finite for all $0<\veps<\sqrt{\lambda}$.
 Since the intervals $I_{n,\veps}$ are nested in $\veps$, the point $\xi_n\in I_{n,\veps}$ for any
 $0<\veps\le\veps_0$. In particular, $I_{n,\veps}\subset \{x: |x-\xi_n|\le R_\veps\}$ for all
 $0<\veps\le\veps_0$ by definition of $R_\veps$. Moreover, again since the sets
 $I_{n,\veps}$ are nested, $I_{n,\veps}\subset I_{n,\veps_0}\subset \{x: |x-\xi_n|\le R_\veps\}$
 for all $\veps_0<\veps<\sqrt{\lambda}$, too.

 Putting everything together, \eqref{eq:I-n-veps} shows that
 with this choice of $\xi_n$ and $R_\veps$ the bound \eqref{eq:maximing-seq-tight-modulo-shift} holds.
 Thus it is enough to prove \eqref{eq:nested}, \eqref{eq:finite-length}, and \eqref{eq:I-n-veps}:
 Let
 \beq
  \begin{split}
    a_{n,\veps} &:= \min\big(z\in\Z:\, \sum_{x< z} |f_n(x)|^2 \ge \frac{\veps^2}{2}\big)\\
    b_{n,\veps} &:= \max\big(z\in\Z:\, \sum_{x > z} |f_n(x)|^2 \ge \frac{\veps^2}{2}\big).
  \end{split}
 \eeq
 Both exist and are finite, since $f_n\in l^2(\Z)$. Put $c_{n,\veps}= \min(a_{n,\veps},b_{n,\veps})-1$ and
 $d_{n,\veps}= \max(a_{n,\veps},b_{n,\veps})+1$.
 With this choice \eqref{eq:nested} and \eqref{eq:I-n-veps} certainly hold and we only have to
 check \eqref{eq:finite-length}, which is where Lemma \ref{lem:epsilon-fat-tail} enters.

 Either $a_{n,\veps}> b_{n,\veps}$, in which case $d_{n,\veps}- c_{n\veps}\le 2$, or Lemma
 \ref{lem:epsilon-fat-tail} applies and gives the bound
 \beq\label{eq:rearranged}
    P_1 \frac{\veps^2}{2} -\big(P_\lambda-\varphi(f_n)\big)
    \le \frac{C \|f_n\|_2^4}{ \Big((d_{n,\veps}-c_{n,\veps}-1)^{1/2}-1\Big)^{1/2}}
 \eeq
 where we rearranged \eqref{eq:epsilon-fat-tail-bound} a bit and used the scaling
 $P_1\|f_n\|_2^4 = P_1\lambda^2= P_\lambda$. Since $f_n$ is a maximizing sequence
 for \eqref{eq:variational} we have $\lim_{n\to\infty} \varphi(f_n)= P_\lambda$. Thus taking the limit
 $n\to\infty$ in \eqref{eq:rearranged} gives
 \bdm
   P_1 \frac{\veps^2}{2}
   \le \liminf_{n\to\infty} \frac{C \lambda^2}{ \Big((d_{n,\veps}-c_{n,\veps}-1)^{1/2}-1\Big)^{1/2}}
 \edm
 or
 \bdm
    \limsup_{n\to\infty} (d_{n,\veps}-c_{n,\veps}) \le \Big[\Big( \frac{2C\lambda^2}{P_1\veps^2}\Big)^2+1\Big]^{2} +1 <\infty.
 \edm
 which proves \eqref{eq:finite-length} and hence the Lemma.
\epf
A simple reformulation of Lemma \ref{lem:maximizing-seq-tight} is
\begin{cor} \label{cor:tight-modulo-shifts} Assume that the diffraction profile obeys
the bound \eqref{eq:D-bounded} and  let $\lambda>0$.
Then, given any maximizing sequence $(f_n)_n$ of the variational problem
\eqref{eq:variational}, there exists a sequence of translations $\xi_n$ such that the
translated sequence ${f}_{n,\xi_n}$ with ${f}_{n,\xi_n}(x):= f_n(x-\xi_n)$ is tight, that is,
 \beq\label{eq:tight-modulo-shifts}
   \lim_{L\to\infty} \limsup_{n\to\infty} \sum_{|x|>L} |{f}_{n,\xi_n}(x)|^2 = 0 .
 \eeq
Moreover, this shifted sequence is still a maximizing sequence for the variational problem
\eqref{eq:variational}.
\end{cor}
\bpf
By Lemma \ref{lem:maximizing-seq-tight}, the sequence $f_{n,\xi_n}$ is certainly tight. On the other hand
it is also a maximizing sequence for \ref{eq:variational}, since the time evolution
$T_t= e^{iD(t)\Delta }$ commutes with translation, $T_tf_{n,\xi_n}(x)= T_tf_{n}(x-\xi_n)$ for all $x$,
and hence
 \bdm
    \varphi(f_{n,\xi_n}) = \calQ(f_{n,\xi_n},f_{n,\xi_n},f_{n,\xi_n},f_{n,\xi_n}) =
    \calQ(f_n,f_n,f_n,f_n) = \varphi(f_n)
 \edm
 for all $n\in\N$.
\epf
\begin{thm}[$=$ Theorem \ref{thm:main-existence}]\label{thm:existence}
 Let $\lambda>0$ and assume that the diffraction profile obeys the bound \eqref{eq:D-bounded}.
 Then there is an $f\in l^2(\Z)$ with $\|f\|_2^2=\lambda$ such that
 \bdm
   \varphi(f) = P_\lambda:= \sup_{g: \|g\|_2^2=\lambda} \varphi(g) .
 \edm
 This maximizer $f$ is also a weak solution of the dispersion management equation
 \bdm
   \omega f = Q(f,f,f).
 \edm
 where $\omega=P_\lambda/\lambda>0$ is the Lagrange-multiplier.
\end{thm}
\bpf
Let $\lambda>0$ and $(f_n)_n$ be a maximizing sequence for \eqref{eq:variational}
with $\|f_n\|^2 = \lambda$.
By Corollary \ref{cor:tight-modulo-shifts}, we can, without loss of generality, assume
that this maximizing sequence is already tight in the sense of equation \eqref{eq:tight-modulo-shifts}.

Since the unit ball in $l^2(\Z)$ is weakly compact, there is a subsequence $(f_{n_j})_{j\in\N}$ of
$(f_n)_{n\in\N}$ which converges weakly to some $f\in l^2(\Z)$. From Lemma
\ref{lem:compactness-criterion} we know that $f_{n_j}$ converges even strongly to $f$ in $l^2(\Z)$.
Thus $\|f\|_2^2=\lambda$ and hence $f$ is a good candidate for the maximizer. To conclude that
$f$ is a maximizer for the variational problem, we need to show that
$\varphi(f)= P_\lambda$. Since $\|f\|_2^2=\lambda$ one certainly has
 \bdm
    \varphi(f) \le P_\lambda = \lim_{n\to\infty}\varphi(f_n),
 \edm
so one only needs upper semi-continuity of $\varphi$ at $f$, i.e.,
 \beq\label{eq:upper-semi-continuous}
   \limsup_{n\to\infty} \varphi({f}_{n_j})
   \le \varphi(f) .
 \eeq
By Lemma \ref{lem:continuity} below, the map $f\mapsto \varphi(f)$ is even continuous on $l^2(\Z)$, in
particular, \eqref{eq:upper-semi-continuous} is true which finished the proof that the variational
problem \eqref{eq:variational} has a maximizer.

The proof that the above maximizer is a weak solution of the associated Euler--Lagrange equation
\eqref{eq:DiffMS} is standard in the calculous of variations, we sketch it for the convenience of
the reader: Lemma \ref{lem:derivative} below shows that the derivative of the functional
$\varphi$ at any $f\in l^2(\Z)$ is given by the linear map $D\varphi(f)[h]= 4\re\calQ(h,f,f,f)$.
Similarly, one can check that the derivative of $\psi(f)=\|f\|_2^2= \la f,f\ra$ is given by
$D\psi(f)[h]= 2\re\la h,f\ra$.
Note that although, in our convention for the inner product, the map $h\mapsto \la h,f\ra$ is
anti-linear, the map $h\mapsto \re\la h,f\ra$ is linear. Similarly, one easily checks that
for fixed $f$ the map $h\mapsto \re\calQ(h,f,f,f)$ is linear.

Now let $f$ be any maximizer of the constraint variational problem \eqref{eq:variational} and
$h\in l^2(\Z)$ arbitrary. Define, for any $(s,t)\in\R^2$,
 \begin{align*}
    F(s,t)&:= \varphi(f+ sf +t h) , \\
    G(s,t)&:= \psi(f+ sf +t h ) .
 \end{align*}
Note that
 \begin{align*}
    \nabla F(s,t)&= \left(\begin{array}{c} D\varphi(f+ sf +t h)[f] \\ D\varphi(f+ sf +t h)[h] \end{array} \right)\\
    &= 4\left(\begin{array}{c} \re\calQ(f,f+ sf +t h,f+ sf +t h,f+ sf +t h) \\
                              \re\calQ(h,f+ sf +t h,f+ sf +t h,f+ sf +t h)  \end{array} \right)
 \end{align*}
and
 \bdm
    \nabla G(s,t)= \left(\begin{array}{c} D\psi(f+ sf +t h)[f] \\ D\psi(f+ sf +t h)[h] \end{array} \right)
    = 2\left(\begin{array}{c} \re\la f,f+ sf +t h\ra \\
                              \re \la h,f+ sf +t h\ra  \end{array} \right) .
 \edm
Since $\la f,f\ra = \lambda\not= 0$,
 \bdm
    \nabla G(0,0)= 2\left(\begin{array}{c} \la f,f\ra \\
                              \re \la h,f\ra  \end{array} \right)
 \edm
 is not the zero vector in $\R^2$ and since $\nabla G(s,t)$ depends multi-linearly, in particular
 continuously, on $(s,t)$, the implicit function theorem \cite{Stichartz00} shows that there
 exists an open interval $I\subset\R$ containing $0$ and a differentiable function $\phi$ on $I$
 with $\phi(0)=0$ such that
 \bdm
    \lambda= \|f\|_2^2 = G(0,0) = G(\phi(t),t)
 \edm
 for all $t\in I$. Consider the function $I\ni t\mapsto F(\phi(t),t)$. Since $f$ is a maximizer
 for the constraint variational problem \eqref{eq:variational}, $F(\phi(t),t)$ has a local maximum at
 $t=0$. Hence, using the chain rule,
 \bdm
    0 = \frac{d F(\phi(t),t)}{dt}\Big\vert_{t=0}
        = \nabla F(0,0)\cdot\left(\begin{array}{c}\phi'(0)\\1\end{array}\right)
        = 4 \calQ(f,f,f,f) \phi'(0) + 4\re\calQ(h,f,f,f)
 \edm
 Since $\lambda= G(\phi(t),t)$, the chain rule also yields
 \bdm
    0= \frac{d G(\phi(t),t)}{dt}\Big\vert_{t=0}
        = \nabla G(0,0)\cdot\left(\begin{array}{c}\phi'(0)\\1\end{array}\right)
        = 2 \la f,f\ra \phi'(0) + 2\re \la h,f\ra.
 \edm
 Solving this for $\phi'(0)$ and plugging it back into the expression for the derivative of $F$,
 we see that
 \bdm
    \frac{\calQ(f,f,f,f)}{\la f,f\ra} \re \la h,f\ra = \re\calQ(h,f,f,f).
 \edm
 In other words, with $\omega := \calQ(f,f,f,f)/\la f,f\ra= P_\lambda/\lambda >0$ and $f$ any maximizer of
 \eqref{eq:variational}, we have
 \beq\label{eq:euler-lagrange-real}
     \re(\omega\la h,f\ra) = \re\calQ(h,f,f,f)
 \eeq
 for any $h\in l^2(\Z)$. Replacing $h$ by $ih$ in \eqref{eq:euler-lagrange-real}, one gets
 \beq\label{eq:euler-lagrange-imaginary}
     \im(\omega\la h,f\ra) = \im\calQ(h,f,f,f)
 \eeq
 for all $h\in l^2(\Z)$. \eqref{eq:euler-lagrange-real} and \eqref{eq:euler-lagrange-imaginary} together
 show
 \bdm
    \omega \la h,f\ra = \calQ(h,f,f,f)
 \edm
 for any $h\in l^2(\Z)$, that is, $f$ is a weak solution of the diffraction management equation
 \eqref{eq:DiffMS}.
\epf
In the proof of Theorem \ref{thm:existence}, we needed the following two Lemmata.
\begin{lem}\label{lem:continuity}
The map $f\mapsto \varphi(f)= \calQ(f,f,f,f)$ is locally Lipshitz continuous on $l^2(\Z)$.
\end{lem}
\bpf
 Given $f,g$, one has
 \beq
 \begin{split}
   \calQ(f,f,f,f) - \calQ(g,g,g,g)
   &=  \int_0^1 \|T_tf\|_4^4 - \|T_tg\|_4^4  \, dt \\
   &= \int_0^1
            \sum_{j=0}^3
                \|T_tf\|_4^{3-j}
                    \big( \|T_tf\|_4 - \|T_tg\|_4 \big)
                \|T_tg\|_4^{j}
        \, dt \, .\\
 \end{split}
 \eeq
 The above together with the triangle inequality, the bound $\|h\|_4\le \|h\|_2$ for all $h\in l^2(\Z)$,
 see Lemma \ref{lem:p-q-norm-bound}, and the unicity of $T_t$ on $l^2(\Z)$ gives
\beq
 \begin{split}
   \big|\calQ(f,f,f,f) - \calQ(g,g,g,g) \big|
    &\le \int_0^1
            \sum_{j=0}^3
                \|T_tf\|_2^{3-j}
                    \big| \|T_tf\|_2 - \|T_tg\|_2 \big|
                \|T_tg\|_2^{j}
        \, dt \\
    &\le \int_0^1
            \sum_{j=0}^3
                \|T_tf\|_2^{3-j}
                    \|T_t(f-g)\|_2
                \|T_tg\|_2^{j}
        \, dt \\
    &\le \sum_{j=0}^3
            \|f\|_2^{3-j} \|f-g\|_2 \|g\|_2^{j} \\
    &\le 4\max(1,\|f\|_2^3, \|g\|_2^3) \|f-g\|_2 .
 \end{split}
 \eeq
\epf
We need one more technical result, about the differentiability of the non-linear functional
$\varphi$.
\begin{lem}\label{lem:derivative}
 The functional $\varphi$ is continuously differentiable with derivative
 $D\varphi(f)[h]= 4\re\calQ(h,f,f,f)$
\end{lem}
\bpf
Using the multi-linearity of $\calQ$, one can check that for any $f,h\in l^2(\Z)$
 \begin{align}
    \varphi(f+h)
    &= \varphi(f)+ \calQ(f,f,f,h)+\calQ(f,f,h,f) + \calQ(f,h,f,f) + \calQ(h,f,f,f)
        + O(\|h\|_2^2) \nonumber \\
    &= \varphi(f) + 4\re\calQ(h,f,f,f) + O(\|h\|_2^2)
 \end{align}
 where in the term $O(\|h\|_2^2)$ we gathered expressions of the form $\calQ(h,f,f,h)$, and
 $\calQ(h,f,f,f+h)$, and similar, which by Corollary \ref{cor:boundedness} are bounded by
 $C\|h\|_2^2$. This shows that $\varphi$ is differentiable with derivative
 $D\varphi(f)[h]= 4\re\calQ(h,f,f,f)$.
 Moreover,
 \bdm
 \begin{split}
    D\varphi(f)[h] - D\varphi(g)[h]
    &= 4\re\big(\calQ(h,f,f,f) - \calQ(h,g,g,g)\big) \\
    & = 4\re\big(\calQ(h,f-g,f,f) + \calQ(h,g,f-g,f) + \calQ(h,g,g,f-g) \big)  .
 \end{split}
 \edm
 Hence using the bound from Corollary \ref{cor:boundedness} again, we see
 \beq
  \sup_{\|h\|_2\le 1} \big| D\varphi(f)[h] - D\varphi(g)[h] \big|
  \le 4(\|f\|_2^2 + \|f\|_2\|g\|_2 + \|g\|_2^2) \|f-g\|_2
 \eeq
 which shows that the derivative $D\varphi$ is even locally Lipshitz continuous.
\epf

%

\smallskip\noindent
\textbf{Acknowledgment:} It is  a pleasure to thank Vadim Zharnitsky
for introducing us to the dispersion management technique.
Young-Ran Lee thanks the Departments of  Mathematics of the University of
Alabama at Birmingham and the University of Illinois at Urbana-Champaign
for their warm hospitality during her visit when most of this work was done.
Young-Ran Lee was partially supported by the Korean Science and Engineering
Foundation(KOSEF) grant funded by the Korean government(MOST)
(No.R01-2007-000-11307-0).

\vspace{5mm}

\renewcommand{\thesection}{\arabic{chapter}.\arabic{section}}
\renewcommand{\theequation}{\arabic{chapter}.\arabic{section}.\arabic{equation}}
\renewcommand{\thethm}{\arabic{chapter}.\arabic{section}.\arabic{thm}}

\def\cprime{$'$}

\end{document}